\documentclass[aps,pre,reprint, amsmath, amssymb,superscriptaddress]{revtex4-1}

\usepackage{morefloats}
\usepackage{bm}
\newcommand{\beq}{\begin{equation}}
\newcommand{\eeq}{\end{equation}}

\usepackage[retainorgcmds]{IEEEtrantools}
\usepackage{graphicx,tikz,placeins}
\usepackage{mathrsfs}
\usepackage{amsmath,amssymb,amsfonts,physics}
\usepackage{color}
\usepackage{float}
\usepackage{times,txfonts}
\usepackage{nicefrac}
\usepackage{verbatim}

\usepackage[colorlinks=true,linkcolor=blue,urlcolor=blue,citecolor=blue,pdfusetitle]{hyperref}
\usepackage{physics}
\tikzset{d/.style={minimum width=7pt,inner sep=0pt,circle,fill=black}}
\usepackage{soul}

\begin{document}
\title{Thermodynamics  of a minimal  collective heat engine: Comparison between engine designs}
%\title{Thermodynamics and performance of a minimal  collective heat engine: Collisional
%approach and distinct  setups}
\author{Felipe Hawthorne}
\affiliation{Universidade de São Paulo,
Instituto de Física,
Rua do Matão, 1371, 05508-090
São Paulo, SP, Brazil}
\date{\today}
\author{B. Cleuren}
\affiliation{UHasselt, Faculty of Sciences, Theory Lab, Agoralaan, 3590 Diepenbeek, Belgium}
\date{\today}
\author{Carlos E. Fiore}
\affiliation{Universidade de São Paulo,
Instituto de Física,
Rua do Matão, 1371, 05508-090
São Paulo, SP, Brazil}
\date{\today}

\begin{abstract} 
Collective effects have attracted remarkable recent interest, not only for their presence in several systems in nature but also for the possibility of being used for the construction of efficient engine setups.
Notwithstanding, little is known about the influence of the engine design and most studies are restricted to the simplest cases (e.g. simultaneous contact with two thermal baths),  not necessarily constituting a realistic setup implementation. Aimed at  partially filling this gap, we introduce the  collisional/sequential description  for a minimal
model for collective effects, composed of two
interacting nanomachines placed in contact with a distinct thermal reservoir and nonequilibrium worksource at each stage/stroke.
Thermodynamic quantities are exactly
obtained irrespectively the model
details. Distinct kinds of engines are investigated and the influence of the interaction, temperature, period, and time asymmetry have been undertaken. Results show that a careful design of interaction provides a superior performance than the interactionless case, including optimal power outputs and efficiencies at maximum power greater than known bounds or even  the system presenting efficiencies close to the ideal (Carnot) limit. We also show that the case of the system simultaneously placed in contact with two thermal reservoirs constitutes a particular case of our framework.

%Collective effects have been proposed as alternative %candidates for engine
%performance. However, little is known about the engine design and results
%are restricted to fixed nonequilibrium conditions. We %address such issue by
%considering minimal setup presenting
%collective effects, composed of two coupled (interacting) quantum-dots.
%We propose a sequential (collisonal) description in %which the system is subject to a distinct condition %at each stroke. Thermodynamics quantities are exactly %solvable, irrespective the model parameters. A %detailed analysis and a comparison between engines
%and distinct designs are undertaken.

\end{abstract}

\maketitle

\section{Introduction}

%\FH{Historically, nonequilibrium thermodynamics has tackled its problems through the lens of maximization of power output or efficiency. Given a system and its associated processes, be it purely physical \cite{PhysRevLett.95.190602,seifert2012stochastic,rana2014single,martinez2016brownian}, biological \cite{liepelt1,liepelt2,terai2022biological}, chemical \cite{seader1982thermodynamic}, or even applied physics, such as in quantum technology \cite{Khan_2021}, nonequilibrium thermodynamics has always excelled at showing its efficiency in describing and unraveling modern day problems.}
The construction of nanoscopic steady-state heat engines has attracted a great deal of recent attention in the realm of  stochastic thermodynamics \cite{seifert2012stochastic, van2005thermodynamic, callen1960thermodynamics,moreira2023stochastic,Fritz_2020}, not
only for extending the fundamental concept
of the energy conversion (from
the macroscopic to the nanoscopic scale), but also
because it presents three fundamental differences when
compared with  the equilibrium thermodynamics. First, there is no need for moving parts and pistons since the energy conversion comes from currents of microscopic particles/units.
Second, nanoscopic-engineered setups typically operate 
far from equilibrium and consequently,
its performance is expected to be lower than the ideal case. Third, fluctuations of quantities and
currents can become important in small-scaled systems. The issues above illustrate the search for the protocol as crucial to
ensure its reliability and desired performance.

In the last years, distinct kinds of engines
operating far from equilibrium
have been proposed and investigated \cite{seifert2012stochastic,Santos_2021,zhao2021stochastic,fu2022stochastic,kumari2023stochastic}.
Under a generic point of view, they are grouped out in three categories, stemming from fixed thermodynamic forces \cite{ge2012stochastic,liepelt1,liepelt2,busiello2022hyperaccurate,mamede2023thermodynamics,gatien,forao2023powerful}, from the time-periodic variation of external parameters \cite{proesmans2015efficiency,proesmans2016brownian,proesmans2017underdamped,mamede2021obtaining}   and via sequential/collisional approach \cite{noa2020thermodynamics,noa2021efficient,mamede2022,rosas1,rosas2,barato2017}, in which at each stroke/stage, the system is subjected to a different  condition (held fixed along the stage).
Each one has been considered as a reliable
approach in distinct contexts, such last one encompassing the presence of distinct drivings over each member of the system, a weak coupling between the system with the reservoir, or even for mimicking the environment for quantum systems \cite{bennett1982thermodynamics,sagawa2014thermodynamic,PhysRevLett.108.040401}. 
While most of the above studies are restricted to  setups composed of one unit \cite{rosas1,harunari2021maximal,proesmans2016brownian,proesmans2017underdamped,mamede2022},
the  thermodynamics of systems exhibiting collective effects has received considerable
recent attention as an alternative strategy for improving the system performance.
 Among the distinct
examples, we cite a system of interacting Brownian particles \cite{mamede2021obtaining}, work-to-work transducers
\cite{herpich,herpich2} and heat engines \cite{gatien,forao2023powerful}. All  of them are restricted to cases of
systems operating 
at equal temperatures \cite{herpich,herpich2}, fixed parameters \cite{gatien,forao2023powerful} or sinusoidal drivings \cite{mamede2021obtaining}.

In this contribution, we conciliate the points above by investigating   a minimal model
for collective effects,  formed by two interacting units  placed sequentially
with distinct thermal baths at each stroke. Previous studies have tackled different versions, such as its all-to-all (mean-field) design \cite{forao2023powerful,gatien}
and distinct topology of interactions \cite{mamede2023thermodynamics}, both for  fixed thermodynamic forces and a large number of units. Our study will focus on the opposite case, dealing with a minimal collective effect system composed of two interacting units beyond the fixed forces context. Hence, its simplicity 
constitutes an ideal laboratory for comparing three fundamental aspects
of nanoscopic engines:  the kind of design (sequential versus fixed thermodynamic forces),   distinct approaches for the worksource (not considered previously) and 
 under situations  collective effects can improve the system performance when compared with its interactionless version.
The former goal has been inspired
from previous contributions \cite{rosas1,harunari2021maximal}, whereas
the different worksources addressed here were considered in Refs.~\cite{rosas1,gatien,mamede2023thermodynamics,forao2023powerful}.
 It is worth mentioning that our system shares some similarities with recent studies about a setup
 composed of two interacting quantum dots under the repeated interactions 
\cite{PhysRevE.107.044102,PhysRevResearch.5.023155}.
  Our findings reveal that  collective effects, together with a suited  design of parameters (energy, period, duration of each stage), can significantly enhance the system's performance. 
  Such remarkable improvement can result in optimal power outputs, efficiencies at maximum power greater than known bounds or even efficiencies approaching to the ideal (Carnot) limit.
  As a side result,  our study shows the simultaneous contact with two thermal baths case \cite{gatien} as the ideal limit of fast switching times.

This paper is organized as follows. In Sec. \ref{sec2} the model and
the main expressions for thermodynamic quantities will be presented.
In Secs. \ref{sec4} and \ref{sec3} we shall analyze in detail
two distinct approaches for our engine setup. Conclusions
will be drawn in Sec. \ref{sec5}.
\section{ Model and Thermodynamics}\label{sec2}

Our minimal model  for collective effects is composed of two interacting units sequentially placed into contact with $N$ distinct reservoirs, each one of duration $\tau_\nu-\tau_{\nu-1}$. The 
total time to complete one cycle being $\tau$. At each stroke, occurring between $\tau_{\nu-1}<t\le \tau_{\nu}$, each unit
can be in a lower ($\sigma_k=0$) or upper ($\sigma_k=1$) state, with individual energies $0$ and $\epsilon_\nu$, respectively. The system is connected to the reservoir $\nu$, with temperature  $\beta_{\nu}=1/(k_B T_{\nu})$, and total energy is  given by
\begin{equation}
    {\tilde \epsilon}^{(\nu)}=V_\nu[(1-\sigma_1)\sigma_2+\sigma_1(1-\sigma_2)]+U_\nu\sigma_1\sigma_2+\epsilon_\nu(\sigma_1+\sigma_2),
    \label{energy}
\end{equation}
where  $U_\nu,V_\nu$ correspond
to distinct 
interaction energies, provided they are in the same and different
states, respectively. Throughout this paper, we adopt $k_B =1$.
%Eq.~(\ref{energy}) is conveniently rewritten in the following way $ {\tilde %\epsilon}^{(\nu)}=(U_\nu-2V_\nu)\sigma_1\sigma_2+(\epsilon_\nu+V_\nu)(\sigma_1+%\sigma_2)$. Since  $U_\nu$ and $V_\nu$ have opposite signals, they will
%present similar roles provided $U_\nu$ (PENSAR AQUI). 
In addition, the system
can be subjected to a non-conservative force $F_\nu$. 

After the time duration $\tau_\nu$, it is disconnected and reconnected to the next reservoir with temperature $\beta_{\nu+1}=1/T_{\nu+1}$ and subjected to another set of parameters
$\epsilon_{\nu+1}$, $U_{\nu+1}$, $V_{\nu+1}$ and $F_{\nu+1}$. This
process is then repeated until a complete cycle, after the total time $\tau$.
As in Refs. \cite{gatien,vroylandt2020isometric,mamede2023thermodynamics},
above system dynamics becomes simpler when  characterized by the total particle number $i$ occupying the upper state, assuming the values $i=0,1$ or $2$, according to whether it is empty, having one unit, or having two units, with
energies ${\tilde \epsilon}^{(\nu)}=0,V_\nu+\epsilon_\nu$ and $U_\nu+2\epsilon_\nu$, respectively.  Let $p^{(\nu)}_i(t)$ be the system's
probability at the state $i$ at the time $t$ when it is placed in contact with the $\nu$-th reservoir, governed by the following master equation
\begin{equation}
{\dot p_i}^{(\nu)}(t)=\sum_{j\neq i}J_{ij}^{(\nu)},
%    \frac{\mathrm{d}\textbf{p}^{(\nu)}(t)}{\mathrm{d}t} = \omega^{(\nu)}\textbf{p}^{(\nu)}(t),
    \label{me}
\end{equation}
where  $J_{ij}^{(\nu)}\equiv \omega^{(\nu)}_{ij}p^{(\nu)}_j-\omega^{(\nu)}_{ji}p^{(\nu)}_i$ and $\omega^{(\nu)}_{ij}$ accounts to the transition rate from state $j$ to $i$, satisfying the condition $\sum_i\omega^{(\nu)}_{ij} =0$ for every strokes.

\begin{center}
\begin{figure}
\includegraphics[scale=0.25]{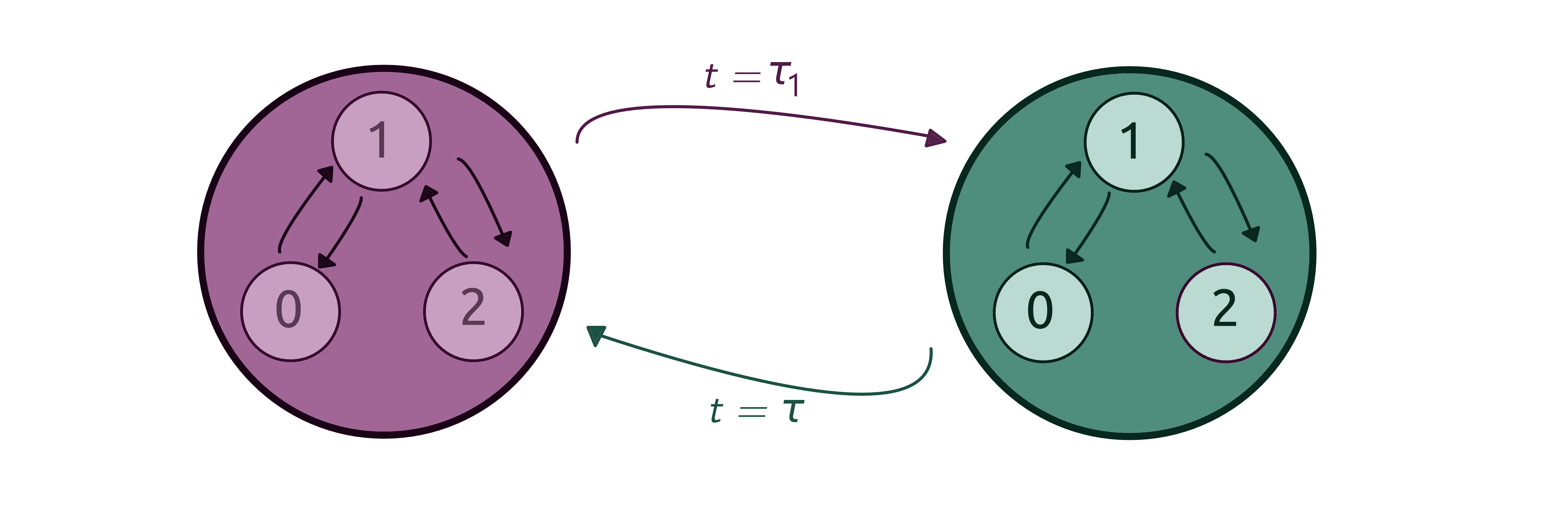}
\caption{Sketch of the setup, composed of two coupled nanomachines, characterized by a three-state system specified by the variable $i$ accounting to the occupation of the upper level.
At each stage (with duration $\tau_1$ and $\tau-\tau_1$) the system is placed in contact with a distinct thermal bath and parameters, specified by the purple and green colors. The stroke change occurs at $t=\tau_1$ and the system returns to its initial state at $t=\tau$.}
\label{f1}
\end{figure}
\end{center}

We shall restrict our analysis to the simplest case $N=2$, as sketched in Fig.~\ref{f1}, in which the time duration of the first and second strokes read $\tau_1$ and $\tau_2=\tau-\tau_1$, respectively. Note that one has the symmetric time operation when $\tau_1=\tau/2$. Given that  the probability distribution is continuous, it should satisfy the following boundary conditions for $p_i^{(\nu)}\left(t\right)$ (for all $i=0,1$ and $2$):
%\begin{subequations}
    \begin{equation}
        p_i^{(1)}\left(\tau_1\right) = p_i^{(2)}\left(\tau_1\right),\qquad
        p_i^{(1)}\left(0\right) = p_i^{(2)}\left(\tau\right).
        \label{pc2}
    \end{equation}
%\end{subequations}
%It is worth pointing out that despite the des framework is valid for an arbitrary number of stages, we shall %focus on $N=2$ case, allowing us to derive explicit expressions for
%thermodynamic quantities.

By resorting to the eigendecomposition of Eq.~(\ref{me}) along with the periodic boundary conditions
given by Eq.~(\ref{pc2}), it is possible to obtain the expression
for the probability component $p_i^{(\nu)}(t)$ at the $\nu$-th stage:
\begin{equation}
 p_i^{(\nu)}(t)   = p_i^{(eq,\nu)} +\sum_{j=1}^{2}e^{\lambda^{(\nu)}_j[t-\tau_{i-1}]}\Gamma^{(\nu)}_{j}\textbf{p}^{(\nu)}(\tau_{i-1}),
\end{equation}
where $\textbf{p}^{(eq,\nu)}$ is the  probability vector associated
with $\lambda_0=0$, $\lambda^{(\nu)}_{j}<0$ is the $j$-th non-zero eigenvalue, $\Gamma^{(\nu)}_{j} = \psi^{(\nu)}_{j}\phi^{(\nu)}_{j}$ is the matrix associated with the product of the $j$-th right and left eigenvectors
and $\textbf{p}^{(\nu)}_l$ is the initial condition vector at the start of each stroke, obtained from the above boundary conditions. Despite being exact, expressions for
$\textbf{p}^{(\nu)}_l$
are quite cumbersome. In Appendix \ref{apa}, we list them for the particular case $\tau_1=\tau/2$.

Once the probability distribution, all thermodynamic quantities can be obtained. By integrating Eq.~(\ref{me}) over a complete cycle and by summing them,
one has that $\sum_{j\neq i}{\bar J}^{(1)}_{ij}=-\sum_{j\neq i}{\bar J}^{(2)}_{ij}$, where ${\overline J}^{(1)}_{ij}=\int_{0}^{\tau_1}J^{(1)}_{ij}dt/\tau$, ${\overline J}^{(2)}_{ij}=\int_{\tau_1}^{\tau}J^{(2)}_{ij}dt/\tau$
and 
Eq.~(\ref{pc2}) was used. 
%\begin{eqnarray}
%   \frac{1}{\tau}\{ p^{(1)}_i\left(\tau_1\right)-p^{(1)}_i\left(0\right)\}&=&\frac{1}{\tau}\sum_{j\neq i}{\overline J}^{(1)}_{ij},\nonumber\\
% \frac{1}{\tau}\{p^{(2)}_i\left(\tau\right)-p^{(2)}_i\left(\tau_1\right)\}&=&\frac{1}{\tau}\sum_{j\neq i}{\overline J}^{(2)}_{ij},
%\label{inte}
%\end{eqnarray}
%where ${\overline J}^{(1)}_{ij}=\int_{0}^{\tau_1}J^{(1)}_{ij}dt$ %and a similar
%expression for ${\overline J}^{(2)}_{ij}$ is straightforwardly obtained by integrating over the second stage. By summing both of %them, it follows that  
At each time,
only transitions  $i\rightarrow i\pm 1$ are allowed, implying that  a transition of type
 $0\leftrightarrow 2$ is forbidden and hence the system presents only two independent fluxes, namely ${\bar J}^{(1)}_{01}$ and ${\bar J}^{(1)}_{21}$, whose expressions are
given by
\begin{subequations}
    \begin{equation}
        {\bar J}^{(1)}_{01} =\frac{1}{\tau} \int_{0}^{\tau_1}\left\{\omega^{(1)}_{01}p^{(1)}_1(t) - \omega^{(1)}_{10} p^{(1)}_0(t)\right\}dt
    \end{equation}
     \begin{equation}
        {\bar J}^{(1)}_{21} = \frac{1}{\tau}\int_{0}^{\tau_1}\left\{\omega^{(1)}_{21}p^{(1)}_1(t) - \omega^{(1)}_{12} p^{(1)}_2(t)\right\}dt,
    \end{equation}
    \label{fr2}
\end{subequations}
respectively.
We pause to make a few comments about fluxes ${\bar J}^{(1)}_{ij} $'s. Firstly, in the regime of fast switchings $\tau \rightarrow 0$, where each stroke relaxes infinitely fast to the steady state, each flux ${\bar J}^{(\nu)}_{ij}$ acquires a simpler form listed in
Appendix \ref{apb}. They can be alternatively obtained by assuming the system relaxes “infinitely fast” to the steady state in such a way that
\begin{equation}
 {\bar J}^{(\nu)}_{ij}\rightarrow \frac{1}{2}\left\{\omega_{ij}^{(\nu)}p_j-\omega_{ji}^{(\nu)}p_i  \right\},
\end{equation}
for $j=1$ and $i\in\{0,2\}$, where 
$p_i=p^{(1)}_i+p^{(2)}_i$, whose expressions for $p_i$ are listed in Appendix \ref{apb} and are equivalent to the system simultaneously placed in contact with both thermal baths.
Secondly, in the regime of slow switchings, $\tau\gg1$, 
\begin{equation}
{\bar J}^{(\nu)}_{ij}\rightarrow \frac{(-1)^{(\nu+1)}}{\tau}\left\{p_i^{(eq,1)}-p_i^{(eq,2)} \right\},
\end{equation}
which vanishes as $\tau \rightarrow \infty$, consistent with the case of the system being placed in contact with a single thermal reservoir.

Until here, all analyses have been carried out without any thermodynamic consideration.
For that, we follow the common approach considered in the literature (see, \textit{e.g.}, Refs. \cite{fiorek,seifert2012stochastic,tome2015,gatien,forao2023powerful}) in which  the ratio between transition rates $\omega^{(\nu)}_{ij}$ and $\omega^{(\nu)}_{ji}$ are defined according to the local detailed balance:
\begin{equation}
\ln\frac{\omega^{(\nu)}_{ij}}{\omega^{(\nu)}_{ji}}=-\beta_\nu\left [ \tilde{\epsilon}^{(\nu)}_i-\tilde{\epsilon}^{(\nu)}_j+d^{(\nu)}_{ij}F_\nu \right],
\label{trans}
\end{equation}
where $\tilde{\epsilon}^{(\nu)}_i-\tilde{\epsilon}^{(\nu)}_j$ is the difference between states $i$ and $j$ and $d^{(\nu)}_{ji}F_\nu$
accounts to the influence of a driving force, where element $d^{(\nu)}_{ji}$ satisfies the anti-symmetric property $d^{(\nu)}_{ji}=-d^{(\nu)}_{ij}$.

From Eq.(\ref{trans}) we consider   the entropy production formula 
\begin{equation}
\Pi_{\nu}(t)=\sum_{i<j}J_{ij}^{(\nu)}(t)\ln \frac{\omega^{(\nu)}_{ij}p_j(t)}{\omega^{(\nu)}_{ji}p_i(t)},
\end{equation}
whose integration over a complete cycle, together with the previously mentioned boundary conditions leads
to the standard form ${\bar \sigma}=-\sum_\nu \beta_\nu{\overline {\dot Q}}_{\nu}$, where 
${\overline{\dot Q}}_{\nu}$ is given by
\begin{equation}
{\overline {\dot Q}}_{\nu}=\sum_{i<j}\left [ \tilde{\epsilon}^{(\nu)}_i-\tilde{\epsilon}^{(\nu)}_j+d^{(\nu)}_{ij}F_\nu \right]{\overline J}^{(\nu)}_{ij}.
\label{genh}
\end{equation}
By expressing Eq.~(\ref{genh}) in terms
of fluxes ${\bar J}^{(\nu)}_{01}$ and ${\bar J}^{(\nu)}_{21}$, the exchanged
heat ${\overline {\dot Q}}_{\nu}$ then reads
\begin{equation}
{\overline {\dot Q}}_{\nu}=\left[\left({\tilde \epsilon}^{(\nu)}_{0}-{\tilde\epsilon}^{(\nu)}_{1}+d^{(\nu)}_{01}F_\nu\right){\bar J}^{(\nu)}_{01}+\left({\tilde \epsilon}^{(\nu)}_{2}-{\tilde\epsilon}^{(\nu)}_{1}+d^{(\nu)}_{21} F_\nu\right){\bar J}^{(\nu)}_{21}\right].
\label{heat}
\end{equation}  
Since the system evolves to a nonequilibrium steady state regime  returning
to the initial state after a complete cycle, the first law of thermodynamics establishes that
${\cal P}=-({\overline {\dot Q}}_{1}+{\overline {\dot Q}}_{2})$, and hence the expression for ${\cal P}$ is  given by 
\begin{equation}
\begin{split}
{\cal P}&=-[\sum_{i}\left(\tilde{\epsilon}^{(2)}_i-\tilde{\epsilon}^{(1)}_i\right)\frac{\left(p^{(1)}_i(\tau_1)-p^{(1)}_i(0)\right)}{\tau} \\
&+\sum_{i<j}d^{(1)}_{ij}\left(F_1 {\overline J}^{(1)}_{ij}-F_2 {\overline J}^{(2)}_{ij}\right)],
\end{split}
\label{power}
\end{equation}
where Eqs.~(\ref{pc2}) and (\ref{heat}) were used, together the properties: $d^{(1)}_{ij}=-d^{(2)}_{ij}$,
$d^{(\nu)}_{ij}=-d^{(\nu)}_{ji}$
and ${\overline J}^{(\nu)}_{ij}=-{\overline J}^{(\nu)}_{ji}$. Above equation states
that the power output comes from two worksources: the former, from the time variation of energies (first term) after each
stroke and the latter from non-conservative forces (second term).
By expressing in terms of independent fluxes, Eq.~(\ref{power})
reads
 ${\cal P}=\left[\left({\tilde \epsilon}^{(2)}_{0}-{\tilde\epsilon}^{(2)}_{1}\right)-\left({\tilde \epsilon}^{(1)}_{0}-{\tilde\epsilon}^{(1)}_{1}\right)\right]{\bar J}^{(1)}_{01}+\left[\left({\tilde \epsilon}^{(2)}_{2}-{\tilde\epsilon}^{(2)}_{1}\right)-\left({\tilde \epsilon}^{(1)}_{2}-{\tilde\epsilon}^{(1)}_{1}\right)\right]{\bar J}^{(1)}_{21}-\left(d^{(1)}_{01}{\bar J}^{(1)}_{01}+d^{(1)}_{21}{\bar J}^{(1)}_{21} \right)\left(F_1+F_2\right)$.
 
Finally,  by defining the second stage as the hot reservoir and choosing parameters properly,
an amount of heat extracted from the hot bath 
$ {\overline {\dot Q}}_{2}>0$ can be partially
converted into power output $ {\cal P}<0$ ($ {\overline {\dot Q}}_{2}=-{\cal P}-{\overline {\dot Q}}_{1}$), consistent to the heat engine operation. Conversely, the pump regime is characterized by 
an amount of power  required for delivering heat or particles
from the cold to the hot reservoir, implying that $ {\cal P}=-{\overline {\dot Q}}_{1}-{\overline {\dot Q}}_{2}$ with $ {\cal P}>0$  and ${\overline {\dot Q}}_{2}<0$. 
In both cases, we adopted the  efficiency definition $\eta=-{\cal P}/{\overline {\dot Q}}_{2}$, implying that the former and latter regimes have efficiencies constrained
according to
$0\le \eta< \eta_c$ and $\eta_c< \eta\le \infty$, respectively, where $\eta_c=1-\beta_2/\beta_1$ denotes Carnot efficiency. 
%From now on,
%we shall adopt the $\eta={\hat \eta}/\eta_c$ as a measure of the system efficiency, implying that the former and latter inequalities read $0\le {\hat %\eta}< 1$ and $0\le 1/{\hat \eta}\le 1$, respectively.

Despite the simplicity, the model presents a great number of parameters $(\beta_\nu,\epsilon_\nu,V_\nu,U_\nu,F_\nu)$ and one of our main goals
is to draw a comparison with previous results \cite{gatien,vroylandt2020isometric} in which solely units in distinct states interact with each other. %were taken into account%
For this reason, we shall curb ourselves to the case $U_\nu=0$.
   \section{Distinct  interactions at each stroke}\label{sec4}
%In order to draw a comparison with previous results \cite{gatien,vroylandt2020isometric,mamede2023thermodynamics} we set $U_\nu=0$ along this section.
\subsection{ Main expressions and general findings}\label{sec4a}
Our first approach consists of  building a setup via change of individual and interaction energies at each stroke without non-conservative drivings. Transition rates $\omega^{(\nu)}_{ij}$ follow Eq.~(\ref{trans}) and have been defined in the following form 
\begin{eqnarray}
\omega^{(\nu)}_{10}&=&2\Gamma \exp\{\frac{-\beta_\nu}{2}(V_\nu+\epsilon_{\nu})\},\\
\omega^{(\nu)}_{01}&=&\Gamma \exp\{\frac{-\beta_\nu}{2}(-V_\nu-\epsilon_{\nu})\},\\
\omega^{(\nu)}_{21}&=&\Gamma \exp\{\frac{-\beta_\nu}{2}(-V_\nu+\epsilon_{\nu})\},\qquad {\rm and }\\
\omega^{(\nu)}_{12}&=&2\Gamma \exp\{\frac{-\beta_\nu}{2}(V_\nu-\epsilon_{\nu})\}
\end{eqnarray}
%$\omega^{(\nu)}_{10}=2\Gamma \exp\{\frac{-\beta_\nu}{2}(V_\nu+\epsilon_{\nu})\}$, $\omega^{(\nu)}_{21}=\Gamma \exp\{\frac{-\beta_\nu}{2}(-V_\nu+\epsilon_{\nu})\}$ (analogous for $\omega^{(\nu)}_{01}$ and $\omega^{(\nu)}_{12}$), 
where $V_\nu,\epsilon_\nu$ assume distinct values at each stroke and $\Gamma$ expresses the coupling between the system and the reservoir. From Eq.~(\ref{genh}), the average heat flux at each stroke is given by 
\begin{align}
        \overline{\dot {Q}}_{1} &= -\left[ \bar{J}^{(1)}_{01}(V_1+\epsilon_1)+\bar{J}^{(1)}_{21}(V_1-\epsilon_1)\right],\nonumber\\
        \overline{\dot{Q}}_{2} &=\left[\bar{J}^{(1)}_{01}(V_2+\epsilon_2)+\bar{J}^{(1)}_{21}(V_2-\epsilon_2) \right],
        \label{heat1}
\end{align}
whose steady entropy production  ${\overline \sigma}$ assumes the generic "fluxes times forces" form ${\overline \sigma}=J_1X_1+J_2X_2$, where $J_1=\bar{J}^{(1)}_{01}$
and   $J_2=\bar{J}^{(1)}_{21}$ with $X_1$ and $X_2$ given by
\begin{eqnarray}
X_1&=&\frac{V_1+\epsilon_1}{T_1}-\frac{V_2+\epsilon_2}{T_2},\nonumber\\
X_2&=&\frac{V_1-\epsilon_1}{T_1}-\frac{V_2-\epsilon_2}{T_2}.
\end{eqnarray}
Expressions for the power ${\cal P}$ and system efficiency $\eta$ are given by
\begin{equation}
    {\cal P}= (\epsilon_1 - \epsilon_2)(\bar{J}^{(1)}_{01}-\bar{J}^{(1)}_{21})+(V_1-V_2)(\bar{J}^{(1)}_{01}+\bar{J}^{(1)}_{21}),
    \label{p1}
\end{equation}
and
\begin{equation}
\eta=-\frac{(\epsilon_1-\epsilon_2)(\bar{J}^{(1)}_{01}-\bar{J}^{(1)}_{21})+(V_1-V_2)(\bar{J}^{(1)}_{01}+\bar{J}^{(1)}_{21})}{\epsilon_2(\bar{J}^{(1)}_{01}-\bar{J}^{(1)}_{21})+V_2(\bar{J}^{(1)}_{01}+\bar{J}^{(1)}_{21}) },
\label{e1}
\end{equation}
respectively.
We pause again to make a few comments. First, Eqs.~(\ref{heat1})-(\ref{e1}) are general for the two-stroke case, irrespective of the period, asymmetry  and  model parameters. Second, in the absence of interactions ($V_1\rightarrow 0$ and $V_2\rightarrow 0$), the system becomes equivalent to the interactionless setup investigated in Refs.~\cite{rosas1,rosas2,noa2023thermodynamics} for $\mu_1=\mu_2=0$. In such cases,
Eqs.~(\ref{p1}) and (\ref{e1}) reduce to $\eta_s=1-\epsilon_1/\epsilon_2$ and 
${\cal P}_s=(\epsilon_1-\epsilon_2){\bar J}_s$, respectively, where
${\bar J}_s$ (for $\tau_1=\tau/2$) reads:
\begin{equation}
\begin{split}
{\bar J}_s&=\\
&\frac{\prod_\mu \left\{-1 +\cosh\left[\tau\cosh\left(\frac{\beta_\mu \epsilon_\mu}{2}\right)\right]+\sinh\left[\tau\cosh\left(\frac{\beta_\mu \epsilon_\mu}{2}\right)\right]\right\}}{\left(e^{\beta_1\epsilon_1}-e^{\beta_2
\epsilon_2}\right)^{-1}\prod_{\mu'}\left(1+e^{\beta_{\mu'}\epsilon_{\mu'}}\right)\left[-1 + \cosh\left(\tau X\right)+\sinh\left(\tau X\right)\right]},
\end{split}
    \end{equation}
where $X = \cosh\left(\beta_1\epsilon_1/2\right)+ \cosh\left(\beta_2\epsilon_2 /2\right)$.
Both $\eta_s$ and ${\cal P}_s$ can be related through expression ${\cal P}_s=-\epsilon_2\eta_s {\overline J}_s$ consistent to  heat engine  characterized by ${\bar J}_s>0$ ~(since $\beta_1\epsilon_1>\beta_2\epsilon_2$), ${\cal P}_s<0$, $0\leq\eta_s \leq\eta_c$.
Conversely, the pump is characterized by the other way around of
conditions
${\bar J}_s<0$~(since $\beta_1\epsilon_1<\beta_2\epsilon_2)$, ${\cal P}_s>0$ and $\eta_c<\eta_s\leq1$.
Third, contrasting with the interactionless case, there are two independent fluxes, $\bar{J}^{(1)}_{01}$ and $\bar{J}^{(1)}_{21}$, revealing that the interaction between units gives rise to a much richer behavior
than the single case \cite{rosas1}. 
 Eqs.~(\ref{heat1}) and (\ref{p1})
 impose some constraints on the operation regime.
In particular, the heat engine occurs when both inequalities
$(\epsilon_2 - \epsilon_1)(\bar{J}^{(1)}_{01} - \bar{J}^{(1)}_{21})<(V_1-V_2)(\bar{J}^{(1)}_{01} + \bar{J}^{(1)}_{21})$ and $\epsilon_2(\bar{J}^{(1)}_{21} - \bar{J}^{(1)}_{01})>V_2(\bar{J}^{(1)}_{01} + \bar{J}^{(1)}_{21})$  
are simultaneously satisfied, whereas the pump regime takes
place for opposite inequalities.
Fourth, our system will operate more efficiently than the interactionless case ($\eta>\eta_s$) if $(\epsilon_1V_2-\epsilon_2V_1)(\bar{J}^{(1)}_{01}+\bar{J}^{(1)}_{21})>0$. The ideal regime operation yields when $\bar{J}^{(1)}_{01},\bar{J}^{(1)}_{21}\rightarrow 0$. For $\epsilon_1/\epsilon_2$ or $V_1/V_2$ held fixed, $\eta=\eta_c$ when $\beta_2V_2=\beta_1V_1$ and $\beta_2\epsilon_2=\beta_1\epsilon_1$, respectively, whose efficiency is
%Efficiencies will be independent of temperatures
 %and fluxes if $\epsilon_1/\epsilon_2=V_1/V_2<\beta_2/\beta_1$ 
 given by
 $\eta=1-V_1/V_2$, akin to the interactionless expression. Conversely,  maximum efficiencies $\eta_{ME}<\eta_c$ if the condition $\epsilon_1/\epsilon_2=V_1/V_2=\beta_2/\beta_1$ is not satisfied. 
Fifth and last, the occurrence
of the pump regime implies at the following relation between parameters $(\beta_2\epsilon_2-\beta_1\epsilon_1)(\bar{J}^{(1)}_{21}-\bar{J}^{(1)}_{01})>(\beta_2V_2+\beta_1V_1)(\bar{J}^{(1)}_{21}+\bar{J}^{(1)}_{01})$.  \\Figs.~\ref{fig2},~\ref{fig3} and  Appendix \ref{apc} illustrate all above features.   
\subsection{System behavior and heat maps for equal switching times $\tau_1=\tau/2$}

\begin{figure}[h]

\includegraphics[scale=0.6]{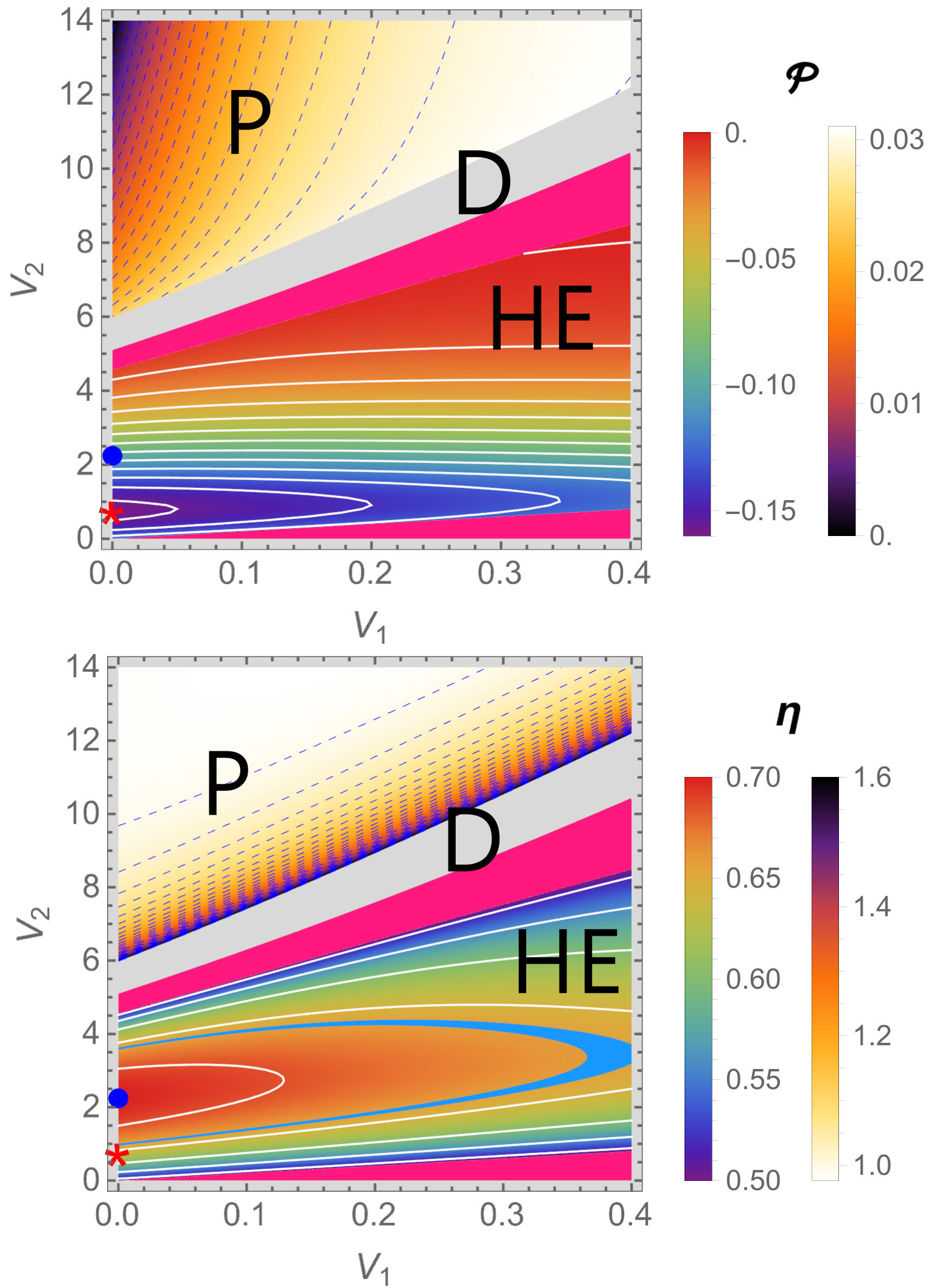}

%\hspace{-1cm}
%\includegraphics[scale=0.62]{fig2a_final.png}

%\hspace{-1cm}
%\includegraphics[scale=0.62]{fig2b_final.png}
     %\caption{The influence interaction parameters over the system performance. Efficiency $\eta$ (left) and power ${\cal P}$ (right) heat maps. Parameters:  $\beta_2=1,\beta_1=10,\tau=1$ and $\epsilon_1/\epsilon_2=0.6$.}
     \caption{The influence of the  interaction parameters over the system performance. Top and bottom panels depict
     the power and efficiency heat maps. The surfaces highlighted by the color pink represent the region in which $\eta \le \eta_s$. Parameters:  $\beta_2=1,\beta_1=10,\tau=1$ and $\epsilon_1/\epsilon_2=0.5$. Symbols  HE (left bars) and P (right bars) 
     correspond to the heat engine and pump regimes, respectively, whereas \textcolor{red}{*} and \textcolor{blue}{$\large\bullet$} attempt to the global maximum of ${\cal P}_{mP}$ and $\eta_{ME}$ in the HE regime. The gray region indicates dud (D) behavior. For this set of parameters $\eta_{ME}<\eta_c$, whereas
     the light blue line in the bottom panel indicates the region in which $\eta_{mP} = \eta_{CA}$.
  }% [The red star, \textcolor{red}{*}, and green square, \textcolor{green}{$\square$} represent the global maxima of the power in the Heat Engine (HE) and Pump (P) regimes, respectively. The gray region indicates que dud (D) behavior. The efficiency, both in the heat engine and pump regimes do not reach Carnot ($0.9$ and $1.11$) respectively.]}
      \label{fig2}
\end{figure}

Once introduced the main expressions, we are now in a position
to depict the system behavior and main results. We chose units in such a way that $\beta_\nu V_\nu$ and $\beta_\nu \epsilon_\nu$ are dimensionless.  The analysis will be carried out for the following
set of parameters: $\beta_1=10,\beta_2=1$, $\tau=1$. %and $k_B=1$.
In order to obtain a first insight into
how the interaction between units influences the system performance,  Fig. \ref{fig2} depicts  
the system performance for $\epsilon_1/\epsilon_2=0.5$, in which
the interactionless case operates as an engine with power and
efficiency given by ${\cal P}_s=-0.1477$ and $\eta_s=0.5$, respectively.

We highlight two remarkable changes coming from the interaction, 
under suitable choices of $V_1(V_2)$ at strokes $v=1(2)$. The former is a broad set of parameters, in which $\eta>\eta_s$
 and ${\cal P}>{\cal P}_s$. The inclusion of interactions
also extends the regime of operation, giving rise
to a pump regime as $V_2$ is raised. Similar results are found
for distinct $\beta_1/\beta_2$'s, although the variation of temperatures can
favor a given operation regime (see e.g. Appendix \ref{apc}).
\begin{figure*}[tp]
\centering
\includegraphics[scale=0.65]{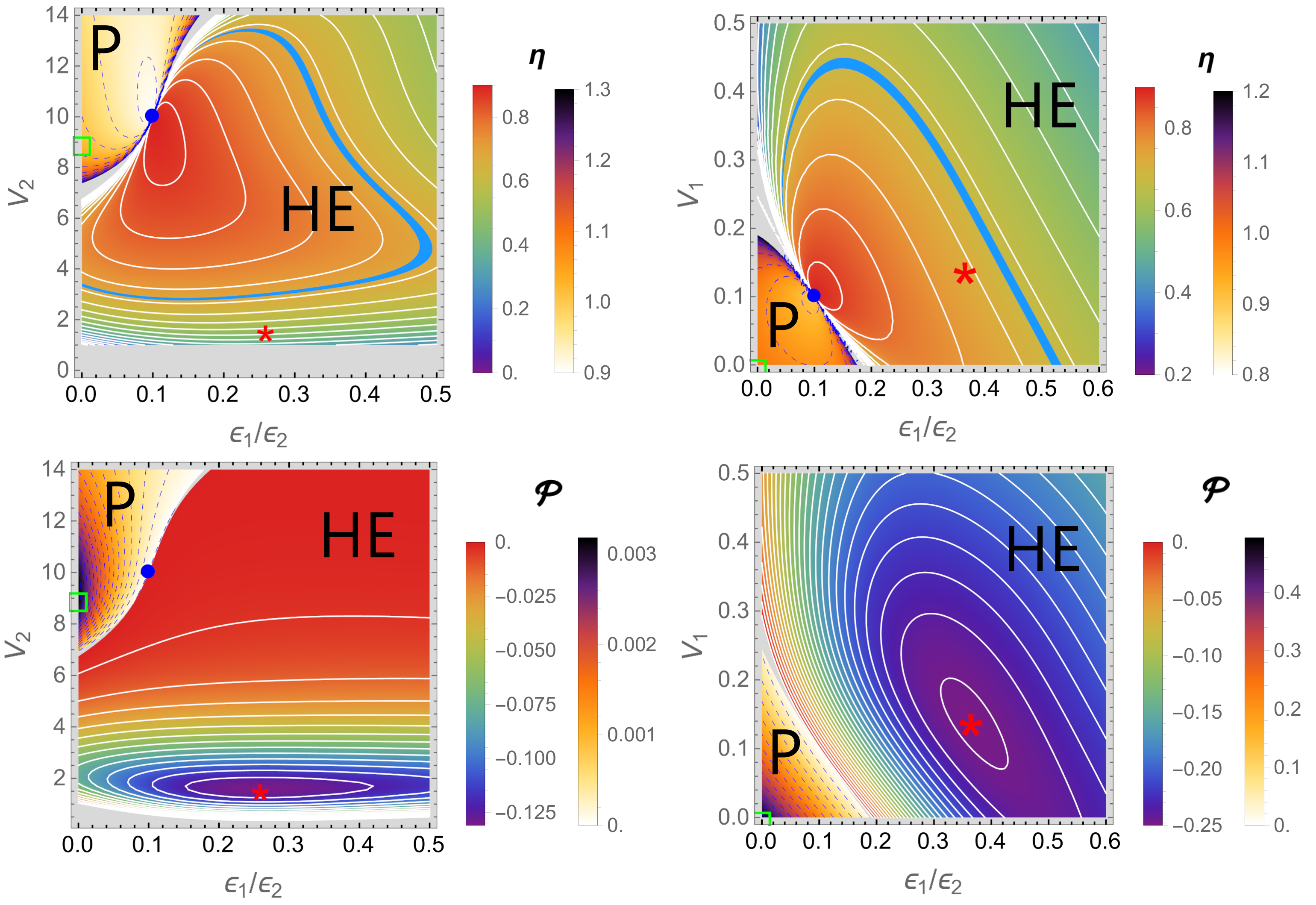}
%\hspace{-2cm}
 %\includegraphics[scale=0.43]{fig3a.png} 
%\begin{subfigure}      
%\hspace{}
%\includegraphics[scale=0.43]{fig3b.png} 
%\end{subfigure}
%\\
%\hspace{-2cm}
%\includegraphics[scale=0.44]{fig3c.png} 
%\begin{subfigure}
%\hspace{}
 %   \includegraphics[scale=0.45]{fig3d.png} 
%\end{subfigure}

        \caption{The influence of individual energies $\epsilon_1/\epsilon_2$ over the system performance. Left and right
     panels depict fixed $V_1$ and $V_2$, respectively, while top and bottom panels show  $\eta$'s and ${\cal P}$'s  heat maps, respectively. Left and right bars denote HE and P regimes, respectively.
    Symbols \textcolor{blue}{$\large\bullet$},   \textcolor{red}{*}
    and \textcolor{green}{$\square$} attempt to Carnot efficiency $\eta_c$, efficiencies at maximum power $\eta_{mP}$ at the heat engine (HE) and pump (P) regimes, respectively. Light blues in top panels indicate the regions in which $\eta_{mP} = \eta_{CA}$.  Parameters:  $\beta_2=1,\beta_1=10,\tau=1$, $V_2=1$ (right) and $V_1=1$ (left).}
      \label{fig3}
\end{figure*}

The interplay between individual $\epsilon_1/\epsilon_2$ and interaction $V_1/V_2$
energies is depicted in Fig. \ref{fig3}, in which $\eta<\eta_s<\eta_c$ for small $V_2$'s.
However, its increase not only extends the heat regime to the region $0<\epsilon_1/\epsilon_2<\beta_2/\beta_1$, in which the interactionless case
operates as a pump,  but also leads to higher efficiencies $\eta>\eta_s$
as $V_2$ increases and a maximum value $\eta_{ME}$ at $V_{2ME}$.
As portrayed in Sec.~\ref{sec4a}, $\eta_{ME}<\eta_c$ for $\epsilon_1\beta_1\neq\epsilon_2\beta_2$ 
and $\eta_{ME}=\eta_c$
at $V_2=V_{2ME}$ when $\epsilon_1\beta_1=\epsilon_2\beta_2$ (e.g. blue $\bullet$ in Fig.~\ref{fig3}) and 
 the interactionless case is efficient in such latter case. 
Similarly to $\eta$, it is possible to find suitable values
of parameters in which ${\cal P}>{\cal P}_s$ 
(from now on meaning the absolute value of ${\cal P}$)
as well as optimize it via a suitable choice of $V_{2mP}$.
However, there is a remarkable difference with respect to $\eta$, the existence of 
an optimal set of  $\epsilon_1/\epsilon_2$ and $V_2$  in which 
(the absolute) ${\cal P}$ is simultaneously maximized (see e.g. symbol $*$ in bottom heat maps).

%{\bf conclusões: fixo $V_1,\epsilon_1/\epsilon_2$ e aumento
%$V_2 \rightarrow$ eficiência aumenta, atinge um valor máximo (Carnot para pequeno %$\epsilon_1/\epsilon_2$) para valores mais altos de $V_2$ e muda o regime de operação. Ela %ultrapassa $\eta_s$ numa
%faixa considerável de $V_2$ à medida que este aumenta. A potência apresenta um máximo %global num range intermediário de $V_2$. O aumento de $\epsilon_1/\epsilon_2$ leva a %conclusões similares, embora a máxima eficiencia diminui um pouco e se move para valores %maiores de $V_2$}
The influence of $V_1$ ($V_2$ held fixed) is remarkably different
from left panels  ($V_1$ held fixed), and
the engine regime and higher efficiencies are constrained to small values of $V_1$'s (consistent with the general findings from Sec.~\ref{sec4a}), hence
pointing us out  that stronger interactions in the second stage are more significant than in the first one (second stage operating as the hot thermal bath).  Also, $\eta>\eta_s$ for a broader set of values of $V_1$ 
as $\epsilon_1/\epsilon_2$ is large.
The behavior of ${\cal P}$ is akin to the previous one and
 presents a maximum at a (small) $V_{1mP}$'s (fixed $\epsilon_1/\epsilon_2$) 
as well as an optimal $\epsilon_1/\epsilon_2$ providing
its simultaneous   maximization.

As a side analysis, we compare efficiencies at maximum power ${\eta_{mP}}$ with Curzon and Ahlborn bound $\eta_\text{CA} = 1- \sqrt{\beta_2/\beta_1}$ \cite{curzon1975efficiency}, which has been verified in distinct systems \cite{novikov1958efficiency, van2005thermodynamic,forao2023powerful}. Despite not constituting a universal result, it provides a powerful guide about the system operation at finite power, which is  more realistic than the ideal case ($\eta=\eta_c$ and ${\cal P}=0$). In all cases,
the interaction among units can also be chosen for providing
efficiencies at maximum power ${\eta}_{mP}>\eta_{CA}$ for a wide range of parameters (see e.g. light blue lines in Figs.~\ref{fig2}-\ref{fig3} in which
$\eta_{mP}=\eta_{CA}$).
Depending on
the parameters the engine is projected, ${\eta}_{mP}<\eta_{CA}$ [Figs.~\ref{fig2} and \ref{fig3} (left panel) ] and ${\eta}_{mP}>\eta_{CA}$ (right panel of Fig.~\ref{fig3})
at the simultaneous maximization of power. 

Summarizing our findings,
 the presence of collective effects between two units
 makes it possible to conveniently choose interaction parameters at each stage, providing higher performances
 than its interactionless counterpart (for the same values
 of individual energies), as well as distinct optimization
 routes, such as the maximization of power and efficiency. Additionally, an extra advantage concerns the possibility of changing the regime operation, from heat engine
 to pump and vice-versa, by changing the interactions at each stroke.
 
\subsection{Influence of period $\tau$ and asymmetric switchings}\label{sec4d}
The influence of period $\tau$ and the inclusion of
a different time duration at each stroke, expressed
by $\kappa=\tau_1/\tau_2\neq 1$ will be considered in this
section. Due to the existence of several distinct parameters, we shall
focus on parameters $\epsilon_1/\epsilon_2=0.6$, $V_1=0.2$, $\beta_1=10$ and $\beta_2=1$.

\begin{figure*}[tp]
\includegraphics[scale=0.7]{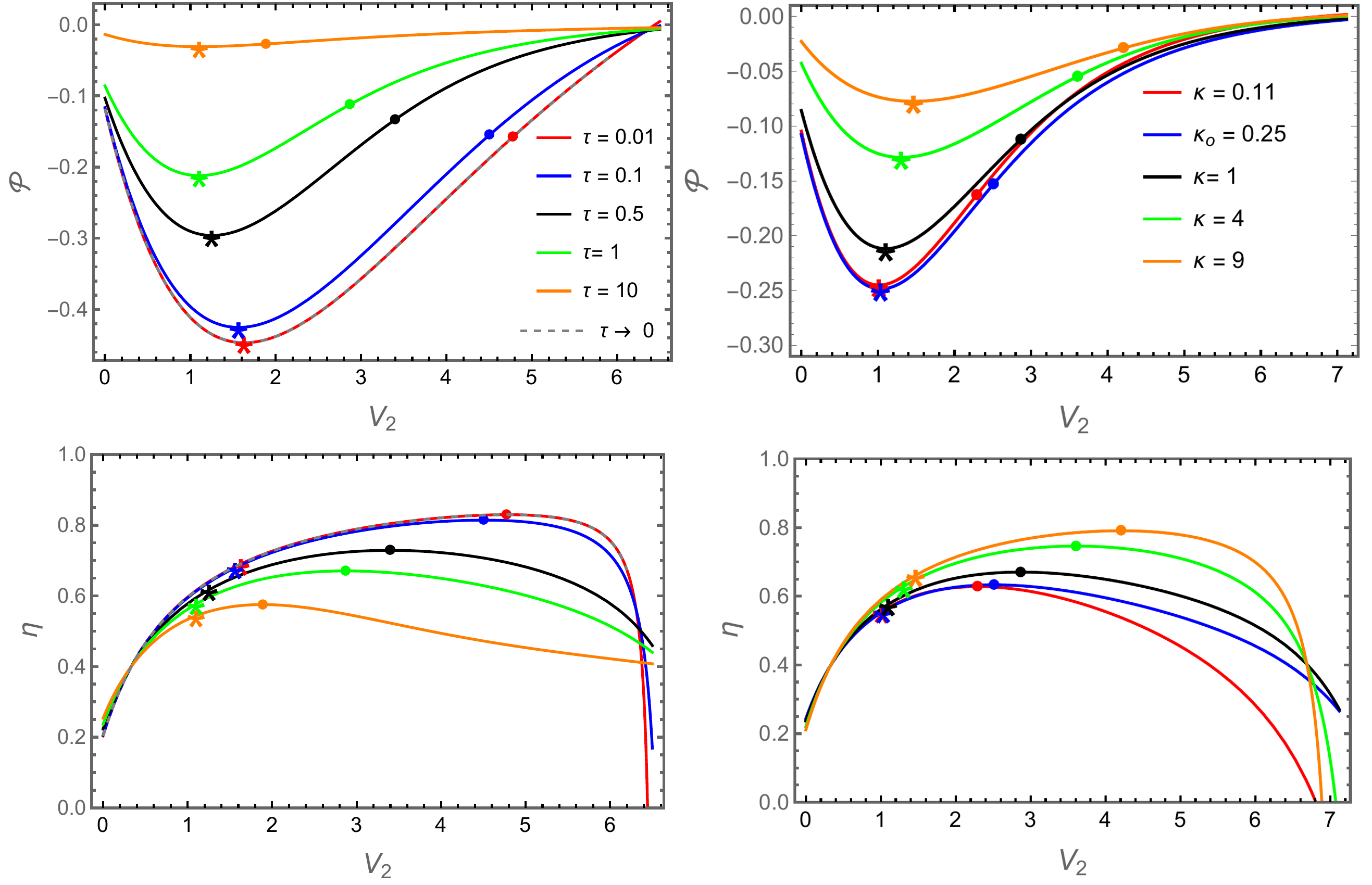}
     \caption{Left and right panels depict the influence
     of period $\tau$ (for symmetric time switchings) and 
     distinct 
     $\kappa$'s (for $\tau=1$), respectively, for ${\cal P}$ (top) and $\eta$ (bottom), respectively. Symbols $*$ and
     $\bullet$ denote associate ${\cal P}_{mP}$'s and $\eta_{ME}$'s, respectively. Parameters: $\beta_1 = 10$, $\beta_2 = 1$, $V_1 = 0.2$, $\epsilon_1/\epsilon_2 = 0.6$.}
      \label{fig4}
\end{figure*}

Although  ${\cal P}_s$ increases as $\tau$ is lowered,
the period plays a less important role in the interactionless case, in part because $\eta_s$ is independent of $J_s$ and $\tau$
\cite{rosas1,noa2023thermodynamics}. On the other hand, the existence of two independent fluxes, as a consequence
of  the interaction between nanomachines makes the influence of $\tau$
more revealing. 
We highlight two aspects regarding the influence of $\tau$, as depicted in the left panels of Fig.~\ref{fig4}. First, it significantly affects the system performance, marking  the increase of both ${\cal P}$ (as the interactionless system) and $\eta$ (unlike the interactionless), with  increasing maximum
    ${\cal P}_{mP}$ and $\eta_{ME}$ at $V_{2mP}$ and $V_{2ME}$, respectively, as $\tau$ is decreased toward the limit $\tau\rightarrow 0$, in which the system becomes 
    equivalent to the (simultaneous) contact with hot and cold thermal baths (see e.g. Appendix \ref{apb}). Second, despite the increase of $\tau$ reduces ${\cal P}$ and $\eta$, it enlarges
    the heat engine operation. Thus, the period can be conveniently
    chosen to obtain a desirable compromise between the system performance 
    (power and efficiency) and the range of the operation regime.

A second aspect to be investigated in this section relies on the inclusion of a distinct
duration of each stage, measured by the asymmetry
$\kappa$. This ingredient  has been revealed to be a powerful ingredient for improving the system power
in the interactionless case
 \cite{harunari2021maximal} or even both ${\cal P}$ and $\eta$ in the
 case of collisional Brownian engines \cite{noa2021efficient}
 and is depicted in the right panels of Fig.~\ref{fig4}. Although
$\eta$ typically increases as $V_2$ raises and $\kappa$ (or $\tau_1$) is reduced, consistent with the system placed in contact with the hot thermal bath
during a larger interval, there is an optimal $\kappa_o$ ensuring
optimal power ${\cal P}_{mP}$. 
Thus, like the interactionless case \cite{harunari2021maximal}, $\kappa$
can be conveniently chosen in order to
increase the power-output and ${\cal P}>{\cal P}_s$.
Since $\eta>\eta_s$ for
a broad range of $V_2$'s, the interaction
offers an extra advantage in which $\kappa$ can be suitably chosen in order to obtain the desired $\eta$ (greater than $\eta_s$)
or even the desired compromise between ${\cal P}$
and $\eta$.

%, respectively, whose 
%thermodynamic quantities are also given by Eqs.~(\ref{heat1}),(\ref{p1}) and (\ref{e1}), but now
%fluxes ${\bar J}^{(1)}_{01}$ and ${\bar J}^{(1)}_{21}$ read    ${\bar J}^{(1)}_{01} = %\int_{0}^{\tau_1}\left\{\omega^{(1)}_{01}p^{(1)}_1(t) - \omega^{(1)}_{10} p^{(1)}_0(t)\right\}dt$ %and 
%   ${\bar J}^{(1)}_{21} = \int_{0}^{\tau_1}\left\{\omega^{(1)}_{21}p^{(1)}_1(t) - \omega^{(1)}_{12} %p^{(2)}_0(t)\right\}dt$, respectively.

%\subsection{Linear regime}
%It is instructive to draw a comparison between the above
%results near the equilibrium regime
%when $X_1=X_2=0$, implying that $V_1=V_2, \epsilon_1=\epsilon_2$ and $T_1=T_2$. For that, we can %resort
% to the linear theory \cite{callen1960thermodynamics}, one has
%the fluxes are proportional to thermodynamic forces $J_1=L_{11}X_1+L_{12}X_2$ and %$J_2=L_{21}X_1+L_{22}X_2$,
%where $L_{ij}$ corresponds to associate Onsager coefficient
%evaluated by $L_{ij}=(\partial J_i/\partial X_j)$ evaluated
%at $X_1=X_2=0$. Expressions for Onsager coefficients are listed below:

%%%%%%%%%%%%%%%%%%%%%%%%%%%%%%%%%%%%%%%%%%%%%%%%%%%%%%%%%%%%%%%%%%%%%%%%%%%
%%%%%%%%%%%%%%%%%%%%%%%%%%%%%%%%%%%%%%%%%%%%%%%%%%%%%%%%%%%%%%%%%%%%%%%%%%%
\section{Collisional machine under non conservative drivings}\label{sec3}
%%%%%%%%%%%%%%%%%%%%%%%%%%%%%%%%%%%%%%%%%%%%%%%%%%%%%%%%%%%%%%%%%%%%%%%%%%%
\subsection{Main expressions and heat maps}\label{sec3a}

\begin{figure*}[tp]
\centering
\includegraphics[scale=0.5]{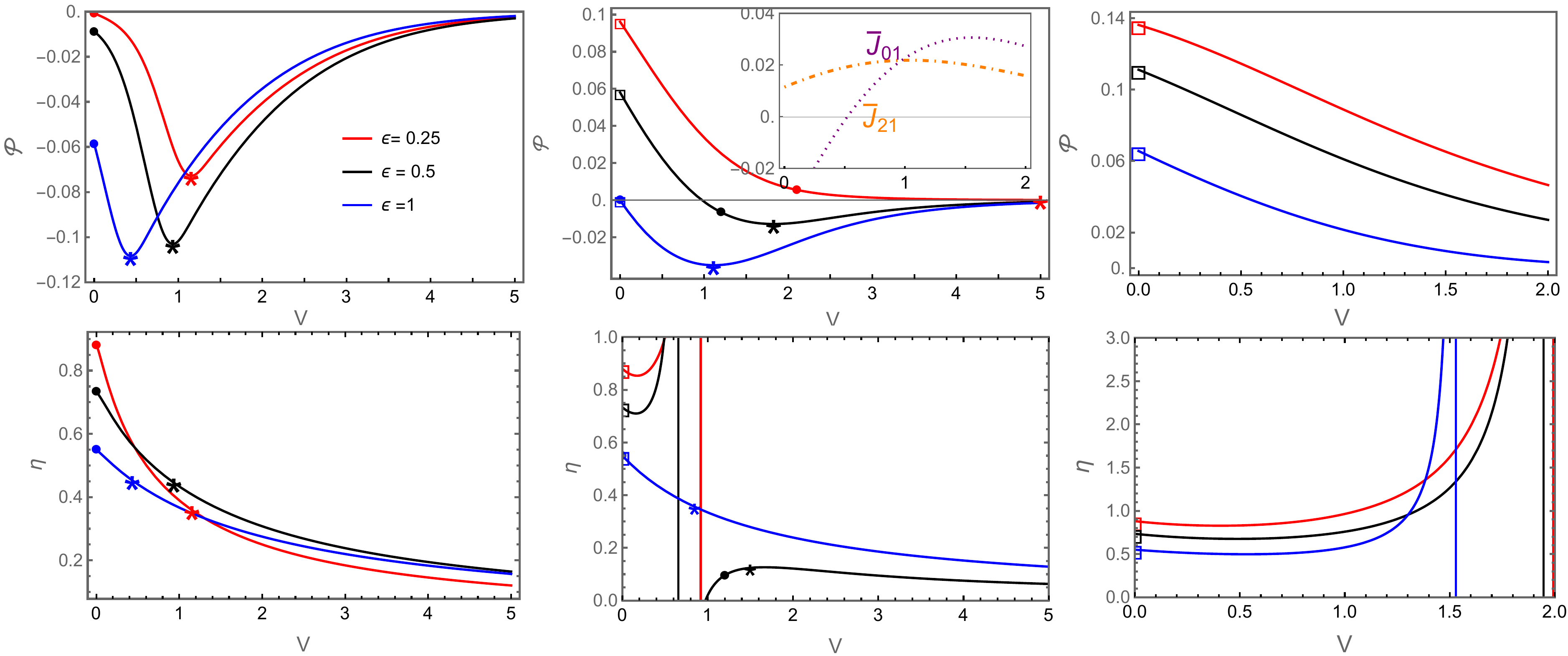}
     \caption{Depiction of   power ${\cal P}$ (top)
     and efficiency ${\hat \eta}$ (bottom) versus $V$ for distinct $\beta_1$'s. Parameters: $\beta_2 =1,E_a = 1, F_2=1,F_1=0.1$ and $\tau =1$ and
      $\beta_1=10$ (left), $\beta_1=20/9$ (middle) and $\beta_1=3/2$ (right).  Stars and squares denote the location of ${\cal P}_{mP}$'s for heat engine and pump, respectively. Circles denote the location of maximum efficiencies for the engine ($0\le \eta<\eta_c$ regime.}
      \label{fig5}
\end{figure*}
Our second approach encompasses a worksource coming from a non-conservative driving,  introduced  by means of 
a bias in order to benefit certain transitions.  By following the ideas of Refs.\cite{gatien,herpich,forao2023powerful}, 
  transitions of type $i\rightarrow i+1$ ($i\rightarrow i-1$)  are favored according to whether the system is placed in contact with the cold (hot) thermal baths, leading to  an incremental  ${\cal P}$
reading $F_\nu$, whereas
the remainig parameters ($V$ and $\epsilon$) are held fixed.  
Our study relies on investigating two important aspects:  the role of drivings at
each stroke and its relationship with $V$, temperatures $\beta_1/\beta_2$ and 
the influence of period $\tau$.
Transition rates $\omega^{(\nu)}_{ij}$ follow Eq.~(\ref{trans}) and are listed below
\begin{eqnarray}
\omega^{(\nu)}_{10}&=&2\Gamma \exp\{\frac{-\beta_\nu}{2}(E_a+V+\epsilon+(-1)^\nu F_\nu)\}\\
\omega^{(\nu)}_{01}&=&\Gamma \exp\{\frac{-\beta_\nu}{2}(E_a-V-\epsilon-(-1)^\nu F_\nu)\}\\
\omega^{(\nu)}_{21}&=&\Gamma \exp\{\frac{-\beta_\nu}{2}(E_a+\epsilon-V+(-1)^\nu F_\nu)\}\\
\omega^{(\nu)}_{12}&=&2\Gamma \exp\{\frac{-\beta_\nu}{2}(E_a-\epsilon+V-(-1)^\nu F_\nu)\},
\end{eqnarray}
%$\omega^{(\nu)}_{10}=2\Gamma \exp\{\frac{-\beta_\nu}{2}(E_a+V+\epsilon+(-1)^\nu F_\nu)\}$, $\omega^{(\nu)}_{21}=\Gamma \exp\{\frac{-\beta_\nu}{2}(E_a+\epsilon-V+(-1)^\nu F_\nu)\}$, $\omega^{(\nu)}_{01}=2\Gamma \exp\{\frac{-\beta_\nu}{2}(E_a-V-\epsilon-(-1)^\nu F_\nu)\}$ and $\omega^{(\nu)}_{12}=\Gamma \exp\{\frac{-\beta_\nu}{2}(E_a-\epsilon+V-(-1)^\nu F_\nu)\}$ 
where $F_\nu$ assumes distinct values at each stroke. Parameter $E_a$ attempts to an activation energy and it will be  included  in order to draw a comparison with previous results \cite{gatien,mamede2023thermodynamics}. Although our main findings are independent of $E_a$, its inclusion makes the heat engine regime more pronounced.   From now  on, we shall set $E_a=1$ in all analyses.  
%From Eq.~(\ref{trans}), one defines
%the  system dynamics in via the following
%transition rates
%$\omega^{(\nu)}_{i,i\pm1}=\Gamma e^{-\frac{\beta_\nu}{2}[\Delta E\mp(-1)^\nu %F_\nu]}$,
%where $\Delta E=\pm \epsilon+{\tilde V}$, where
%${\tilde V}=V$ for transitions $0\rightarrow 1$
%and $2\rightarrow 1$ and $-V$ the other way around.

From Eqs.~(\ref{heat}) and (\ref{power}) and
by taking $V_1=V_2=V$ and $\epsilon_1=\epsilon_2=\epsilon$, the average power and 
 the heat extracted exchanged with the hot bath are given by the following expressions
\begin{eqnarray}
 {\cal P}&=&-(F_1+F_2)({\bar J}^{(1)}_{01}-{\bar J}^{(1)}_{21}),  \label{driving}\\
 {\overline {\dot Q}}_{2}&=&\left[(V+\epsilon+F_2){\bar J}^{(1)}_{01}+\left(V-\epsilon-F_2\right){\bar J}^{(1)}_{21}\right]\nonumber,
 \end{eqnarray}
whose system entropy production reads ${\bar \sigma}=- \beta_{1}{\overline {\dot Q}}_{1}- \beta_{2}{\overline {\dot Q}}_{2}$ and assumes the bilinear form ${\overline \sigma}=J_1X_1+J_2X_2$, where 
$J_1={\bar J}^{(1)}_{01}$ and $J_2={\bar J}^{(1)}_{21}$ (as in Sec. \ref{sec4}) with thermodynamic forces $X_1$ and $X_2$ given by 
\begin{eqnarray}
X_1&=&\frac{\epsilon+V+F_2}{T_2}-\frac{\epsilon+V-F_1}{T_1},\nonumber\\
X_2&=&\frac{\epsilon-V+F_2}{T_2}-\frac{\epsilon-V-F_1}{T_1}.
\end{eqnarray}

The efficiency is given by the ratio between ${\cal P}$ and ${\overline {\dot Q}}_{2}$ given by
\begin{equation}
\eta=\frac{(F_1+F_2)({\bar J}^{(1)}_{01}-{\bar J}^{(1)}_{21})}{(V+\epsilon+F_2){\bar J}^{(1)}_{01}+\left(V-\epsilon-F_2\right){\bar J}^{(1)}_{21}},
\label{etadriving}
\end{equation}
respectively.  The existence of the heat engine and pump regimes imposes some constraints in the fluxes, implying that in the former case parameters have to be adjusted in such a way that ${\bar J}^{(1)}_{01}>{\bar J}^{(1)}_{21}$ and
$V({\bar J}^{(1)}_{01}+{\bar J}^{(1)}_{21})>(\epsilon+F_2)({\bar J}^{(1)}_{21}-{\bar J}^{(1)}_{01})$, whereas the latter (pump)
implies  opposite inequalities.

\begin{figure}[h]
\centering
\includegraphics[scale=0.6]{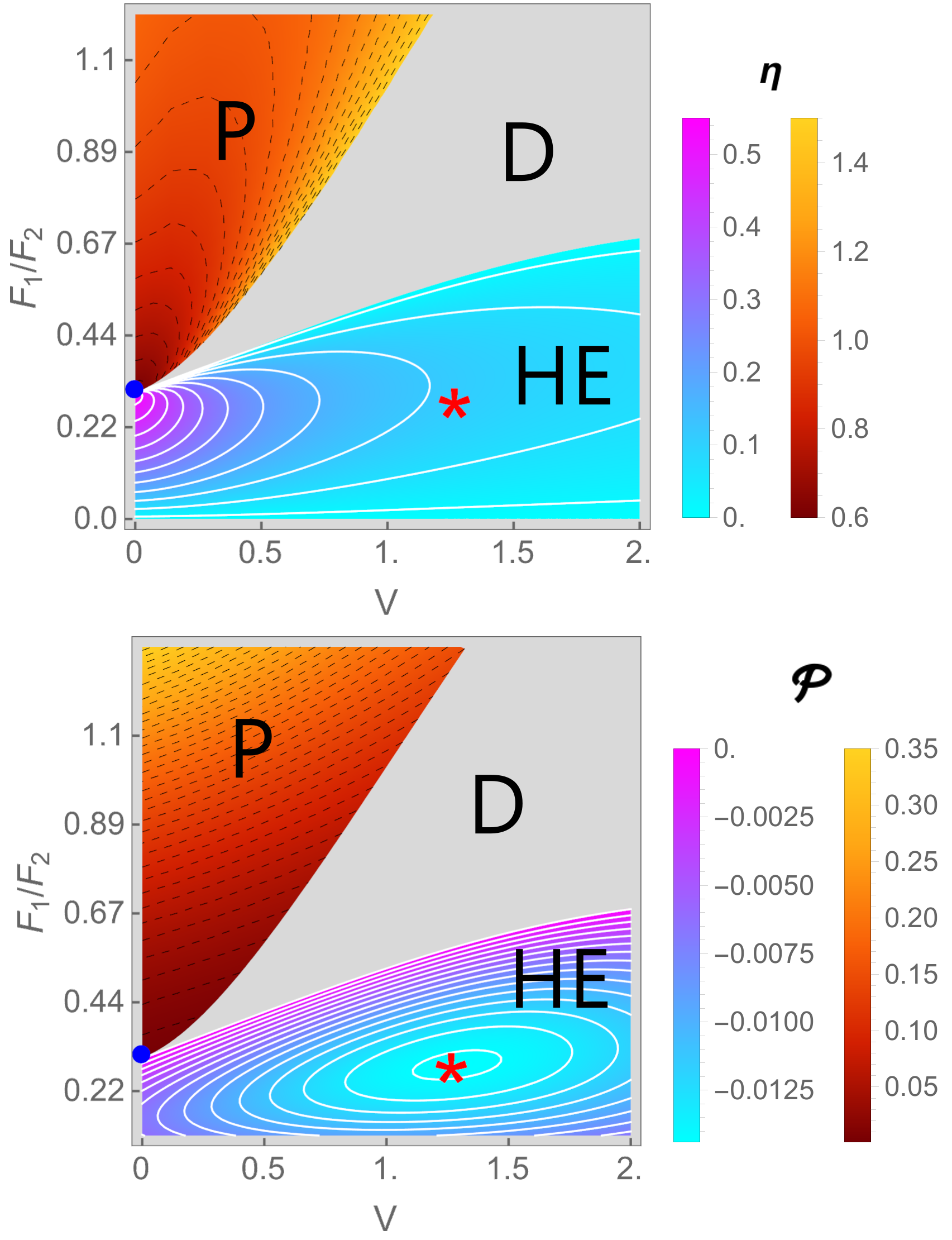}
     \caption{The same as in Fig.\ref{fig2}, but by changing the drivings at each stroke.
     Parameters: $\epsilon=0.5,\beta_2=1,\beta_1=20/9,E_a=1,\tau=1$, $F_2 = 0.45$}
      \label{fig6}
\end{figure}

A first insight about  the influence of drivings is depicted in Figs.~\ref{fig5} for fixed $F_1/F_2$.
Efficiency and power curves exhibit an interesting and rich behavior due to the interplay among parameters $\epsilon, V, \beta_1/\beta_2$ and $\tau$.  
While the heat regime is levered by increasing $\epsilon$ and/or the ratio $\beta_1/\beta_2$ (left and middle panels), the  pump regime is favored for lower values
of $\beta_1/\beta_2$ (middle and right). 
The crossover from the heat to the pump regimes gives rise to an intermediate regime in which
the system operates dudly (see e.g. middle panels). In such
a case, there are
optimal  interactions $V_{mP}$ and $V_{ME}$ marking maximum (absolute) power (${\cal P}_{mP}$) and efficiency ($\eta_{ME}$), respectively. Conversely, only ${\cal P}$ 
can be optimized when the crossover between the above regimes is marked
by the absence of a dud regime  (e.g. left and right panels) and $\eta$ monotonically
decreases upon $V$ being raised.
Fig.~\ref{fig6} extends above findings by depicting heat maps for the efficiency and power 
for distinct ratio $F_1/F_2$ and fixed $\epsilon$. Similarly to systems composed of many interacting units under fixed drivings \cite{forao2023powerful,mamede2023thermodynamics} and results from Sec.~\ref{sec4}, the power ${\cal P}$ presents a simultaneous maximization (concerning both  $V$ and $F_1/F_2$), whereas $\eta$ approaches to the ideal 
regime  $F_2/F_1$ is increased.
However, a difference with respect to previous studies concerns the absence of heat engine as $F_1=F_2$. Unlike Refs.~\cite{gatien,mamede2023thermodynamics,forao2023powerful}, in which the heat engine was investigated for large $N$'s, our minimal setup of $N=2$ interacting units requires a desirable compromise between $F_\nu$'s and parameters  for operating properly as an heat engine.

\begin{figure*}[tp]
\includegraphics[scale=0.6]{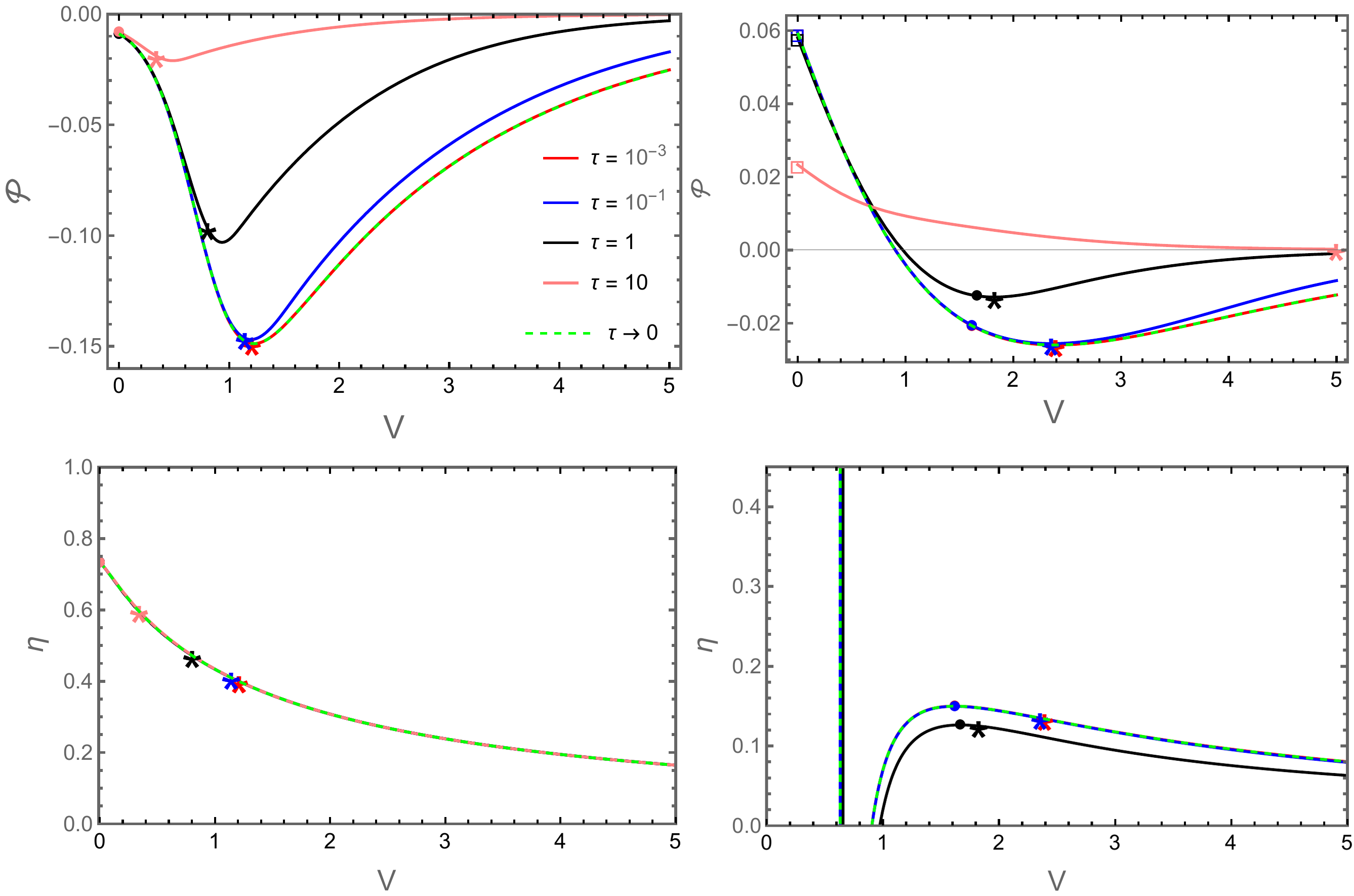}
     \caption{The influence of period $\tau$ over the system performance.
     Depiction of ${\cal P}$ and $ \eta$ versus $V$ for distinct
     $\tau$'s for $\beta_1=10$ (left) and $\beta_1=20/9$ (right). Parameters: $\beta_2=1,E_a=1,\epsilon=0.5,F_2=1,F_1=0.1$.}
      \label{fig7}
\end{figure*}

The influence of period is depicted in Fig.~\ref{fig7} for the same parameters 
from Fig.~\ref{fig6} (left and right panels). In both cases, ${\cal P}$ is strongly
influenced by the period and approaches to the simultaneous contact with 
baths as $\tau \rightarrow 0$, whose expressions are evaluated via Appendix~\ref{apb}. Also, depending on the parameters the engine is
projected (right panels), the increase of $\tau$ changes the regime operation, from heat
engine to pump. In both cases, the behavior of $\eta$ is more revealing  and mildly changes  with
$\tau$. While small differences are almost imperceptible in the left panels, a somewhat
increase of $\eta$ as $\tau$ is lowered is verified. 
 This finding is remarkable, because it may be used for  cconveniently  choosing the period in order to obtain the desirable ${\cal P}$ with a small variation of $\eta$.

\subsection{Asymmetric time switchings}

\begin{figure}[H]
\centering
\includegraphics[scale=0.6]{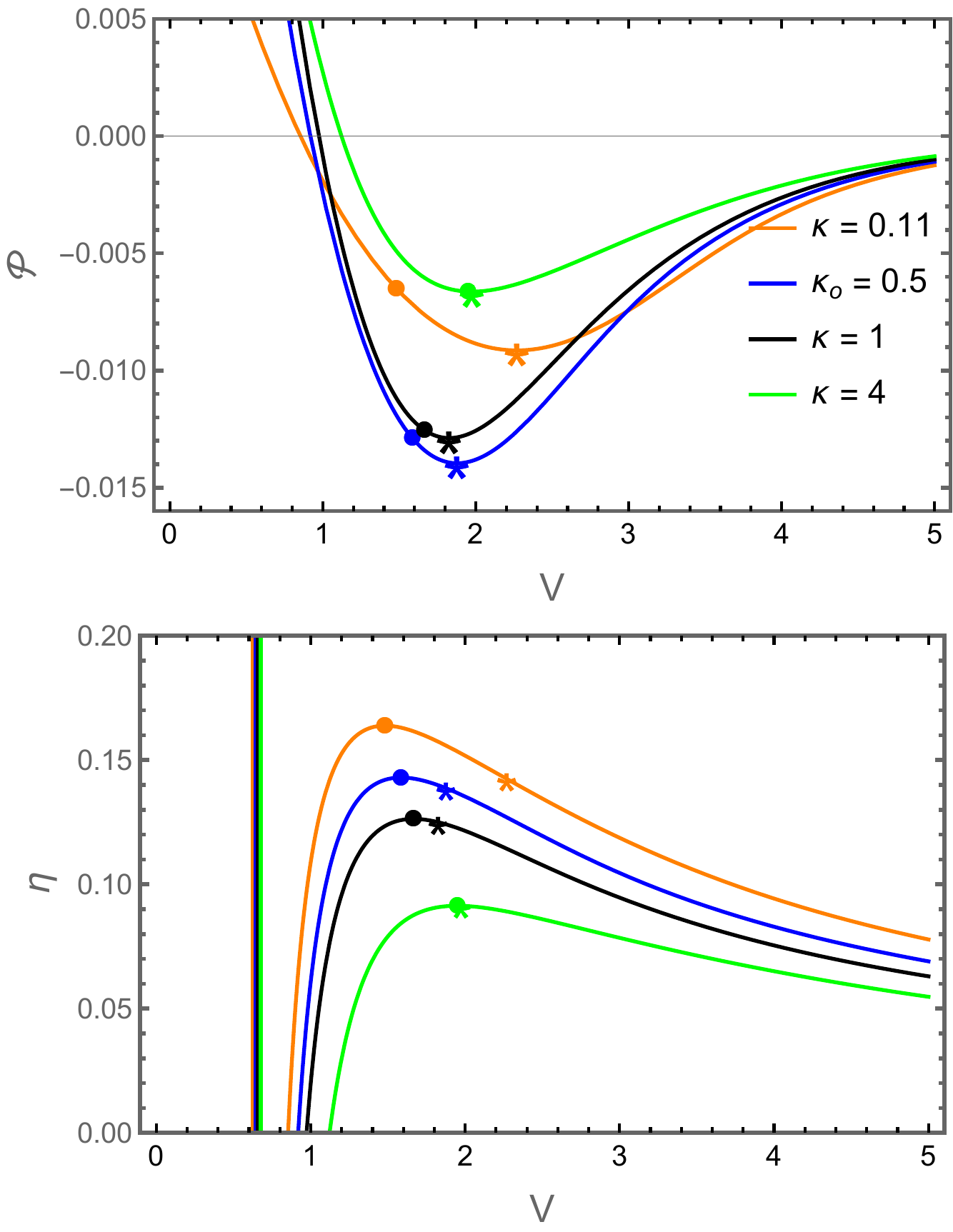}
\caption{Depiction of power ${\cal P}$ and efficiency $\eta$ versus $V$ for different $\kappa$. 
Symbols $\bullet$ and $*$ denote the maximization of efficiency and power, respectively. 
Parameters: $\epsilon=0.5,\beta_2=1,\beta_1=20/9,E_a=1,\tau=1$, $F_1 = 0.1$, $F_2 = 1$.}
      \label{fig8}
\end{figure}

In the last analysis, we investigate  the influence of asymmetric interaction times
in the presence of distinct drivings at each stroke, as shown
in Fig.\ref{fig8}.
In similarity with Fig.~\ref{fig4}, the asymmetry can be conveniently chosen
for enhancing the power and efficiency or even for obtaining a desirable compromise between them. There is an optimal
$\kappa_o$ leading to simultaneous maximization of power while $\eta$ always increases as $\kappa$
is lowered, consistent with the contact with hot bath for a larger amount of the period.
Despite such similarities, the asymmetry seems to be  less pronounced than
in the previous case, and optimal quantities do not deviate significantly from the symmetric ($\kappa=1$) case. A possible reason is that power and efficiency exhibit  a more intricate
dependence  on fluxes and changes of energy parameters (former approach) than on driving
variations [see e.g. Eqs.~(\ref{p1})-(\ref{driving}) and (\ref{e1})-(\ref{etadriving})].

%\begin{figure}
%\includegraphics[scale=0.25]{uneq.pdf}
 %    \caption{The influence of asymmetric time switchings over the system performance.
  %   Depiction of ${\cal P}$ and ${\hat \eta}$ versus $V$ for $\epsilon=0.5$ (top) and $\epsilon=1$ (bottom). Parameters: $\beta_2=1,\beta_1=20/9,\epsilon_a=1,F_2=1,F_1=0.1$.}
   %   \textbf{Vamos manter esta? nessa simulação, não temos interações diferentes entre estágios, somente "V" }\label{fig4}
%\end{figure}

%\subsection{Comparison between distinct engine designs}

%As a means of validation for both the relevance and validity of the results showcased in this work, we present a direct comparison between our findings $-$ obtained through the collisional approach $-$ and the case of simultaneous ($N =2$) heat baths \cite{gatien,rosas1,forao2023powerful}. The exact probability distribution and thermodynamic quantities of interest utilized in this section are derived in Appendix \ref{apb}

%A reasonable comparison between both models can be made in the asymptotic region of small periods, $\tau \ll 1$, when the reservoir switch is fast enough that the system can be understood as in contact with both reservoirs.  

\section{Conclusions}\label{sec5}
Nanoscopic engines operating via collective operation have attracted
considerable attention and posed as potential candidates
for the construction of reliable setups.   However, given that most
studies are restricted to fixed thermodynamic forces,
 little is known about how its construction influences the performance.
The present study aimed to fill partially this gap by investigating the thermodynamic quantities of a minimalist
collective model placed sequentially with distinct thermal baths at each stage. Distinct aspects have been addressed, such as different worksources, the role of interactions,  the period and the time   durations of each stroke. 
Results indicate that our minimal approach, together with a suitable choice of parameters,  not only can boost the system performance, providing  optimal power outputs and efficiencies greater than its interactionless case, but also guide the operation regime, including distinct heat engine and pump regimes. Although the ideal regime   $\tau \rightarrow 0$ provides
higher performances than for finite $\tau$'s, the present contribution sheds light on how the interplay between
interaction and individual parameters, together a suitable  tuning of the interaction time  can 
optimize both power and efficiency as much as possible under more a realistic context (finite $\tau$). 
Another remarkable finding concerns that 
 the case of the system simultaneously placed in
contact with two thermal reservoirs, 
previously investigated in various works \cite{gatien,vroylandt2020isometric,mamede2021obtaining}, constitutes a particular case of our framework for fast switchings. 
As future extensions of our paper, it might be interesting to extend our sequential framework to setups composed of a larger number of nanomachines as well as drawing a comparison among their interactions.

\section{Acknowledgments}
This study was supported by the Special Research Fund
(BOF) of Hasselt University under Grant No. BOF23BL14.
We acknowledge the financial support from CAPES and FAPESP under grants 88887.816488/2023-00 and  2021/03372-2, respectively. The financial support from CNPq is also acknowledged.

\onecolumngrid
\appendix

\section{Obtaining the exact solution for the boundary conditions}\label{apa}
As described in the main text,
by resorting to the eigendecomposition of the evolution matrix, together above boundary conditions, we arrive at the following expression
f or the probability component $p_i^{(\nu)}(t)$ at the $\nu$-th stroke:
\begin{equation}
 p_i^{(\nu)}(t)   = p_i^{(eq,\nu)} +\sum_{j=1}^{2}e^{\lambda^{(\nu)}_j[t-(\nu-1)\frac{\tau}{2}]}\Gamma^{(\nu)}_{j}\textbf{p}^{(\nu)}((\nu-1)\frac{\tau}{2}),
\end{equation}
where $\textbf{p}^{(eq,\nu)}$ is the stationary state probability associated
with $\lambda_0=0$ and $\lambda^{(\nu)}_{j}$ is the $j$-th non-zero eigenvalue and $\Gamma^{(\nu)}_{j} = \psi^{(\nu)}_{j}\phi^{(\nu)}_{j}$ is the matrix associated with the product of the $j$-th right and left eigenvectors
and $\textbf{p}^{(\nu)}((\nu-1)\frac{\tau}{2})$ is the vector at each stroke given by
%\begin{widetext}
%\begin{equation}
 %   \textbf{p}^{(1)}_l =\frac{\sum_{\mu,\mu'}e^{(\lambda^{(1)}_{\mu}+\lambda^{(2)}_{\mu'})\tau/2}\left(\Tr{\Gamma^{(1)}_\mu\Gamma^{(2)}_{\mu'}}-\Gamma^{(1)}_\mu\Gamma^{(2)}_{\mu'}\right)\left(\textbf{p}^{(eq,2)}+e^{\lambda^{(2)}_{\mu'+1}\tau/2}\Gamma^{(2)}_{\mu'+1}\textbf{p}^{(eq,1)}\right)-\left(\prod_{\mu,\eta}e^{\lambda^{(\eta)}_\mu\tau/2} + e^{\omega^{(2)}\tau/2}\right)\textbf{p}^{(eq,1)}}{-1-\prod_{\mu,\eta}e^{\lambda^{(\eta)}_{\mu}\tau/2}+\sum_{\mu,{\mu' }}e^{(\lambda^{(1)}_{\mu}+\lambda^{(2)}_{\mu'})\tau/2}\Tr{\Gamma^{(1)}_{\mu}\Gamma^{(2)}_{\mu'}}}\\
 %\label{p01f}
 %\end{equation}
 %\begin{equation}
  %  \textbf{p}^{(2)}_l=\frac{\sum_{\mu,\mu'}e^{(\lambda^{(1)}_{\mu}+\lambda^{(2)}_{\mu'})\tau/2}\left(\Tr{\Gamma^{(2)}_\mu\Gamma^{(1)}_{\mu'}}-\Gamma^{(2)}_\mu\Gamma^{(1)}_{\mu'}\right)\left(\textbf{p}^{(eq,1)}+e^{\lambda^{(1)}_{\mu'+1}\tau/2}\Gamma^{(1)}_{\mu'+1}\textbf{p}^{(eq,2)}\right)-\left(\prod_{\mu,\eta}e^{\lambda^{(\eta)}_\mu\tau/2} + e^{\omega^{(1)}\tau/2}\right)\textbf{p}^{(eq,2)}}{-1-\prod_{\mu,\eta}e^{\lambda^{(\eta)}_{\mu}\tau/2}+\sum_{\mu,{\mu' }}e^{(\lambda^{(1)}_{\mu}+\lambda^{(2)}_{\mu'})\tau/2}\Tr{\Gamma^{(1)}_{\mu}\Gamma^{(2)}_{\mu'}}}\\
 %\label{p02f}
 %\end{equation}

\begin{equation}
\textbf{p}^{(1)}(0) =\frac{-\left[e^{\tau\omega^{(1)}/2}+e^{\tau\left(\lambda^{(2)}_{1}+\lambda^{(2)}_{2}+\lambda^{(1)}_1+\lambda^{(1)}_2\right)/2}\right]\textbf{p}^{(eq,1)}+\sum\limits_{\nu,\mu,\mu'}e^{\tau\left(\lambda^{(2)}_{1}+\lambda^{(2)}_{2}+\lambda^{(1)}_\nu\right)/2}\Delta^{(1)}_{\mu,\mu'}\left[\Gamma^{(2)}_{\mu'+1}\textbf{p}^{(eq,1)}+e^{-\tau \lambda^{(2)}_{\mu'+1}/2}\textbf{p}^{(eq,2)}\right]}{2\left[\left(e^{\tau\lambda^{(1)}_1/2}-e^{\tau\lambda^{(1)}_2/2}\right)\left(e^{\tau\lambda^{(2)}_1/2}-e^{\tau\lambda^{(2)}_2/2}\right)\Tr{\Gamma^{(1)}_2\Gamma^{(2)}_2} - \left(e^{\tau\left(\lambda^{(1)}_2+\lambda^{(2)}_1\right)/2}-1\right)\left(e^{\tau\left(\lambda^{(1)}_1+\lambda^{(2)}_2\right)/2}-1\right)\right]}
\label{p01f}
\end{equation}

\begin{equation}
\textbf{p}^{(2)}\left(\frac{\tau}{2}\right) =\frac{-\left[e^{\tau\omega^{(1)}/2}+e^{\tau\left(\lambda^{(1)}_{1}+\lambda^{(1)}_{2}+\lambda^{(2)}_1+\lambda^{(2)}_2\right)/2}\right]\textbf{p}^{(eq,2)}+\sum\limits_{\nu,\mu,\mu'}e^{\tau\left(\lambda^{(1)}_{1}+\lambda^{(1)}_{2}+\lambda^{(2)}_\nu\right)/2}\Delta^{(2)}_{\mu,\mu'}\left[\Gamma^{(1)}_{\mu'+1}\textbf{p}^{(eq,2)}+e^{-\tau \lambda^{(1)}_{\mu'+1}/2}\textbf{p}^{(eq,1)}\right]}{2\left[\left(e^{\tau\lambda^{(1)}_1/2}-e^{\tau\lambda^{(1)}_2/2}\right)\left(e^{\tau\lambda^{(2)}_1/2}-e^{\tau\lambda^{(2)}_2/2}\right)\Tr{\Gamma^{(1)}_2\Gamma^{(2)}_2} - \left(e^{\tau\left(\lambda^{(1)}_2+\lambda^{(2)}_1\right)/2}-1\right)\left(e^{\tau\left(\lambda^{(1)}_1+\lambda^{(2)}_2\right)/2}-1\right)\right]}
\label{p02f}
\end{equation}

where
\begin{align*}
    \Delta^{(\nu)}_{\mu,\mu'} =\Tr{\Gamma^{(\nu+1)}_{\mu}\Gamma^{(\nu)}_{\mu'}} -\Gamma^{(\nu+1)}_{\mu}\Gamma^{(\nu)}_{\mu'}
\end{align*}
%\end{subequations}}
%\end{widetext}

\section{The fast time switchings $\tau \rightarrow 0$ and  the two reservoirs case}\label{apb}
In the regime of fast switching dynamics, $\tau \rightarrow 0$, one gets the following expressions for fluxes
\begin{equation}
    \lim_{\tau \to 0} \bar{J}^{(1)}_{01} =\frac{1}{2Z}\left(\omega^{(1)}_{01}\omega^{(2)}_{10}-\omega^{(1)}_{10}\omega^{(2)}_{01}\right)\left(\omega^{(1)}_{12}+\omega^{(2)}_{12}\right),
    \label{b1a}
\end{equation}
and
\begin{equation}
    \lim_{\tau \to 0} \bar{J}^{(1)}_{21} =\frac{1}{2Z}\left(\omega^{(1)}_{21}\omega^{(2)}_{12}-\omega^{(1)}_{12}\omega^{(2)}_{21}\right)\left(\omega^{(1)}_{10}+\omega^{(2)}_{10}\right),
    \label{b1b}
\end{equation}
where $Z=\left(\omega^{(1)}_{01}+\omega^{(2)}_{01}\right) \left(\omega^{(1)}_{12}+\omega^{(2)}_{12}\right)+\left(\omega^{(1)}_{10}+\omega^{(2)}_{10}\right)\left(\omega^{(1)}_{12}+\omega^{(2)}_{12}\right)+\left(\omega^{(1)}_{10}+\omega^{(2)}_{10}\right)
    \left(\omega^{(1)}_{21}+\omega^{(2)}_{21}\right)$.
The above expressions can be understood from the fact  the system relaxes “infinitely fast” to its steady state
at each stroke, allowing to rewrite Eq.(\ref{me}) in the following form
${\dot p^{(\nu)}_i}(t)=\sum_{j\neq i}\{ \omega^{(\nu)}_{ji}p_i(t)-\omega^{(\nu)}_{ij}p_j(t)\}$, where $p_i(t)=p^{(1)}_i(t)+p^{(2)}_i(t)$. Thus, 
the fully dynamics is described by ${\dot p_i}(t)=\sum_{j\neq i}\{ \Omega_{ji}p_i(t)-\Omega_{ij}p_j(t)\}$,
where $\Omega_{ij}=\omega^{(1)}_{ij}+\omega^{(2)}_{ij}$, which is equivalent to the
simultaneous contact with both thermal reservoirs.
A second way of understanding such a limit comes from the time integration of Eq.~(\ref{me}) over
each stage by taking into account the boundary conditions from Eq.~(\ref{pc2}). In such cases, the steady state regime
is given by the following relations $(\omega^{(1)}_{01}+\omega^{(2)}_{01})p_1-(\omega^{(1)}_{10}+\omega^{(2)}_{10})p_0=0$ and $(\omega^{(1)}_{20}+\omega^{(2)}_{20})p_0+(\omega^{(1)}_{12}+\omega^{(2)}_{12})p_2-(\omega^{(1)}_{01}+\omega^{(2)}_{01}+\omega^{(1)}_{21}+\omega^{(2)}_{21})p_1=0$. By solving
above system of linear equations, together with the condition $p_0+p_1+p_2=1$, one finds the following
expressions for the probabilities:
\begin{eqnarray}
    p_0&=&\frac{1}{Z}\left(\omega^{(1)}_{01}+\omega^{(2)}_{01}\right) \left(\omega^{(1)}_{12}+\omega^{(2)}_{12}\right),\nonumber\\
    p_1&=&\frac{1}{Z}\left(\omega^{(1)}_{10}+\omega^{(2)}_{10}\right)
    \left(\omega^{(1)}_{12}+\omega^{(2)}_{12}\right),\\
     p_2&=&\frac{1}{Z}\left(\omega^{(1)}_{10}+\omega^{(2)}_{10}\right)
    \left(\omega^{(1)}_{21}+\omega^{(2)}_{21}\right).\\
    \label{alltoall}
\end{eqnarray}
    It is worth mentioning that $p_i$'s can be alternatively obtained via the spanning tree method for $N=2$. From $p_i$'s, fluxes are promptly obtained, providing
    the same results as Eq.~(\ref{b1a}) and (\ref{b1b}).
    Thermodynamic quantities are straightforwardly evaluated, whose main expressions 
    for ${\cal P}$,
    $\overline{{\dot Q}_{2}}$ and $\eta$ and have been shown along the main text.
    
    We close this section by  pointing out  above expressions are general and hold valid in both Secs. \ref{sec4} and \ref{sec3} when $\tau\rightarrow 0$.
%    As a last1 comment, we 
%    $\langle{\cal P}\rangle=(F_1+F_2)(J^{(1)}_{21}-J^{(1)}_{01})$
%    and , where $J^{(\nu)}_{ij}=\omega^{(\nu)}_{ij}p_j-%\omega^{(\nu)}_{ji}p_i$ and $\langle{\dot Q}}_{2}\rangle=(V+\epsilon+F_2) J^{(1)}_{01}+\left(V-\epsilon-F_2\right) J^{(1)}_{21}$. 

\section{Global phase diagram for distinct interactions at each stroke }\label{apc}
In this section, we depict the system phase diagram  (top panel) built from inequalities,
$(\epsilon_2 - \epsilon_1)(\bar{J}^{(1)}_{01} - \bar{J}^{(1)}_{21})<(V_1-V_2)(\bar{J}^{(1)}_{01} + \bar{J}^{(1)}_{21})$ and $\epsilon_2(\bar{J}^{(1)}_{21} - \bar{J}^{(1)}_{01})>V_2(\bar{J}^{(1)}_{01} + \bar{J}^{(1)}_{21})$, 
shown in the main text for the heat engine (HE) regime and the other way around for the pump (P).
 In particular, the crossover between
HE and P regimes will be characterized by ideal
efficiency provided $\epsilon_1/\epsilon_2=V_1/V_2=\beta_2/\beta_1$ 
(green symbols). The bottom panels show, for different sets of temperatures, the phase 
diagram $V_1/V_2\times \epsilon_1/\epsilon_2 $. As discussed in the main text, while larger
$\beta_1/\beta_2$ favors the HE regime, its decrease increases the region in which
the system operates as a pump.

\begin{figure}[H]
    \centering
    \includegraphics[scale=0.8]{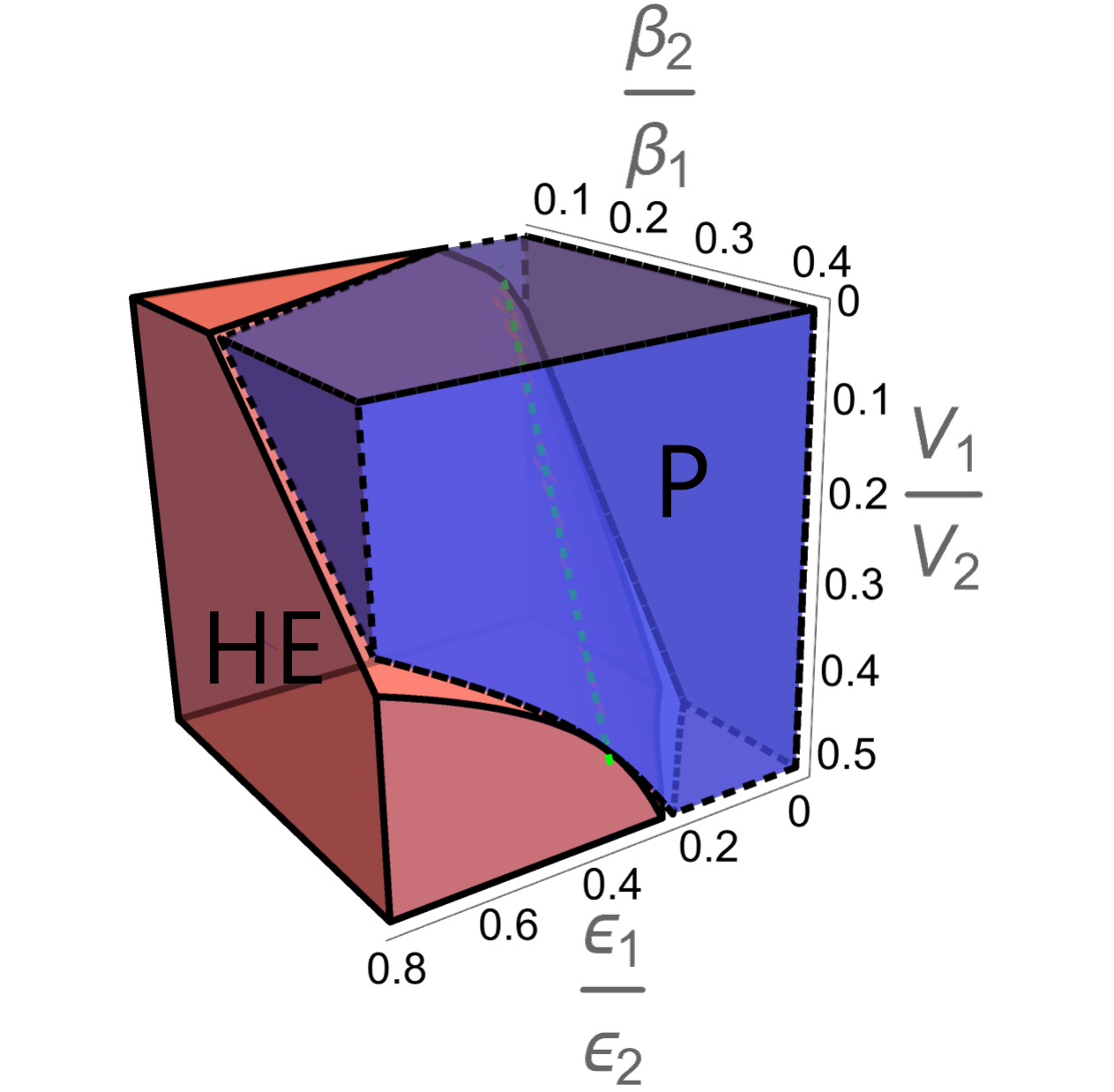}\\
   \includegraphics[scale=0.45]{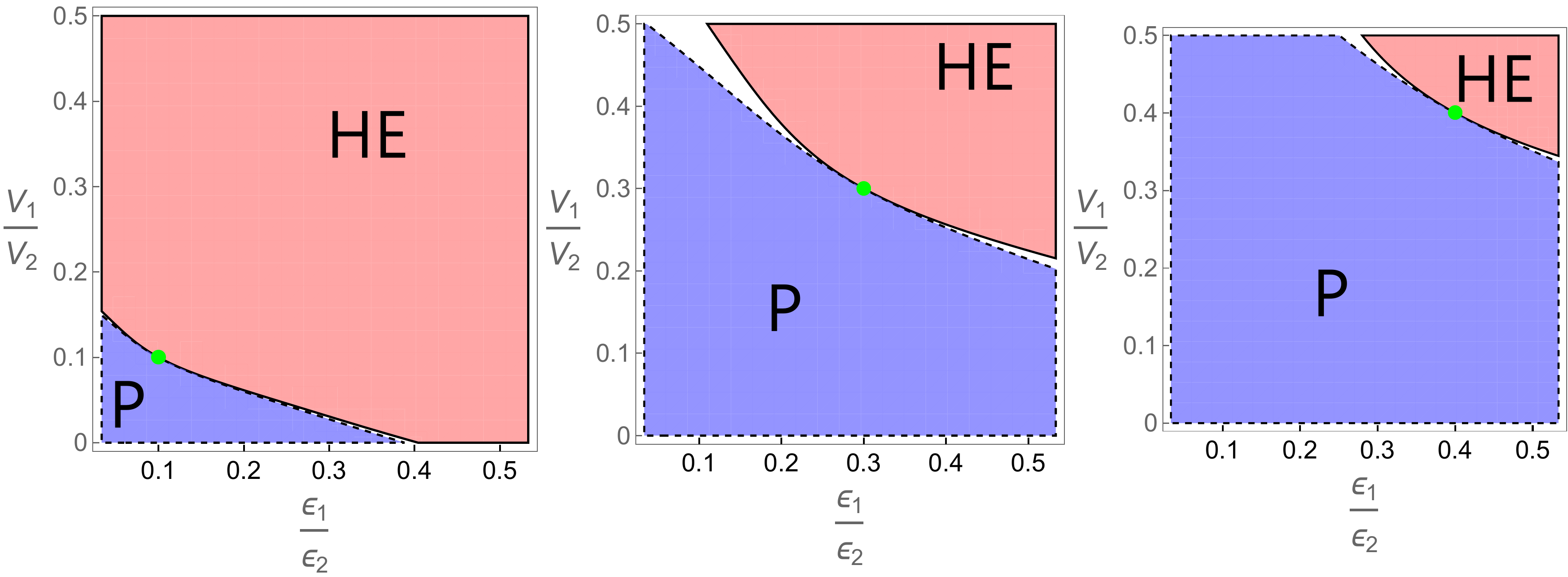}
    \caption{Top panel: The phase diagram $\beta_2/\beta_1\times  V_1/V_2 \times \epsilon_1/\epsilon_2$. The green line represents the points where $\beta_2/\beta_1 = V_1/V_2 = \epsilon_1/\epsilon_2$. Bottom panels depict,
    for $\beta_1 =10$ (left), $\beta_1 = 10/3$ (middle) and $\beta_1 = 10/4$ (right), the 
    phase diagrams in the  $ V_1/V_2\times \epsilon_1/\epsilon_2 $ plane. P and HE denote,
    respectively, the pump and heat engine regimes. White region
    attempts to the dud regime, whereas green bullets correspond
    to the ideal efficiency $\eta_c$. }
    %Parameters: $V_2=5,\epsilon_2=3/2$ and $\beta_2=1$.} 
    \label{pd0}
\end{figure}
\twocolumngrid

\bibliography{refs}

\begin{thebibliography}{40}%
\makeatletter
\providecommand \@ifxundefined [1]{%
 \@ifx{#1\undefined}
}%
\providecommand \@ifnum [1]{%
 \ifnum #1\expandafter \@firstoftwo
 \else \expandafter \@secondoftwo
 \fi
}%
\providecommand \@ifx [1]{%
 \ifx #1\expandafter \@firstoftwo
 \else \expandafter \@secondoftwo
 \fi
}%
\providecommand \natexlab [1]{#1}%
\providecommand \enquote  [1]{``#1''}%
\providecommand \bibnamefont  [1]{#1}%
\providecommand \bibfnamefont [1]{#1}%
\providecommand \citenamefont [1]{#1}%
\providecommand \href@noop [0]{\@secondoftwo}%
\providecommand \href [0]{\begingroup \@sanitize@url \@href}%
\providecommand \@href[1]{\@@startlink{#1}\@@href}%
\providecommand \@@href[1]{\endgroup#1\@@endlink}%
\providecommand \@sanitize@url [0]{\catcode `\\12\catcode `\$12\catcode
  `\&12\catcode `\#12\catcode `\^12\catcode `\_12\catcode `\%12\relax}%
\providecommand \@@startlink[1]{}%
\providecommand \@@endlink[0]{}%
\providecommand \url  [0]{\begingroup\@sanitize@url \@url }%
\providecommand \@url [1]{\endgroup\@href {#1}{\urlprefix }}%
\providecommand \urlprefix  [0]{URL }%
\providecommand \Eprint [0]{\href }%
\providecommand \doibase [0]{http://dx.doi.org/}%
\providecommand \selectlanguage [0]{\@gobble}%
\providecommand \bibinfo  [0]{\@secondoftwo}%
\providecommand \bibfield  [0]{\@secondoftwo}%
\providecommand \translation [1]{[#1]}%
\providecommand \BibitemOpen [0]{}%
\providecommand \bibitemStop [0]{}%
\providecommand \bibitemNoStop [0]{.\EOS\space}%
\providecommand \EOS [0]{\spacefactor3000\relax}%
\providecommand \BibitemShut  [1]{\csname bibitem#1\endcsname}%
\let\auto@bib@innerbib\@empty
%</preamble>
\bibitem [{\citenamefont {Seifert}(2012)}]{seifert2012stochastic}%
  \BibitemOpen
  \bibfield  {author} {\bibinfo {author} {\bibfnamefont {U.}~\bibnamefont
  {Seifert}},\ }\href {\doibase https://doi.org/10.1088/0034-4885/75/12/126001}
  {\bibfield  {journal} {\bibinfo  {journal} {Rep. Prog. Phys.}\ }\textbf
  {\bibinfo {volume} {75}},\ \bibinfo {pages} {126001} (\bibinfo {year}
  {2012})}\BibitemShut {NoStop}%
\bibitem [{\citenamefont {Van~den Broeck}(2005)}]{van2005thermodynamic}%
  \BibitemOpen
  \bibfield  {author} {\bibinfo {author} {\bibfnamefont {C.}~\bibnamefont
  {Van~den Broeck}},\ }\href@noop {} {\bibfield  {journal} {\bibinfo  {journal}
  {Physical Review Letters}\ }\textbf {\bibinfo {volume} {95}},\ \bibinfo
  {pages} {190602} (\bibinfo {year} {2005})}\BibitemShut {NoStop}%
\bibitem [{\citenamefont {Callen}(1960)}]{callen1960thermodynamics}%
  \BibitemOpen
  \bibfield  {author} {\bibinfo {author} {\bibfnamefont {H.~B.}\ \bibnamefont
  {Callen}},\ }\href@noop {} {\bibfield  {journal} {\bibinfo  {journal} {John
  Wiley \& Sons, New York}\ } (\bibinfo {year} {1960})}\BibitemShut {NoStop}%
\bibitem [{\citenamefont {Moreira}\ \emph {et~al.}(2023)\citenamefont
  {Moreira}, \citenamefont {Samuelsson},\ and\ \citenamefont
  {Potts}}]{moreira2023stochastic}%
  \BibitemOpen
  \bibfield  {author} {\bibinfo {author} {\bibfnamefont {S.~V.}\ \bibnamefont
  {Moreira}}, \bibinfo {author} {\bibfnamefont {P.}~\bibnamefont {Samuelsson}},
  \ and\ \bibinfo {author} {\bibfnamefont {P.~P.}\ \bibnamefont {Potts}},\
  }\href@noop {} {\enquote {\bibinfo {title} {Stochastic thermodynamics of a
  quantum dot coupled to a finite-size reservoir},}\ } (\bibinfo {year}
  {2023}),\ \Eprint {http://arxiv.org/abs/2307.06679} {arXiv:2307.06679
  [cond-mat.stat-mech]} \BibitemShut {NoStop}%
\bibitem [{\citenamefont {Fritz}\ \emph {et~al.}(2020)\citenamefont {Fritz},
  \citenamefont {Nguyen},\ and\ \citenamefont {Seifert}}]{Fritz_2020}%
  \BibitemOpen
  \bibfield  {author} {\bibinfo {author} {\bibfnamefont {J.~H.}\ \bibnamefont
  {Fritz}}, \bibinfo {author} {\bibfnamefont {B.}~\bibnamefont {Nguyen}}, \
  and\ \bibinfo {author} {\bibfnamefont {U.}~\bibnamefont {Seifert}},\ }\href
  {\doibase 10.1063/5.0006115} {\bibfield  {journal} {\bibinfo  {journal} {The
  Journal of Chemical Physics}\ }\textbf {\bibinfo {volume} {152}} (\bibinfo
  {year} {2020}),\ 10.1063/5.0006115}\BibitemShut {NoStop}%
\bibitem [{\citenamefont {Santos}\ \emph {et~al.}(2021)\citenamefont {Santos},
  \citenamefont {Tacchino}, \citenamefont {Gerace}, \citenamefont {Campisi},\
  and\ \citenamefont {Santos}}]{Santos_2021}%
  \BibitemOpen
  \bibfield  {author} {\bibinfo {author} {\bibfnamefont {T.~F.~F.}\
  \bibnamefont {Santos}}, \bibinfo {author} {\bibfnamefont {F.}~\bibnamefont
  {Tacchino}}, \bibinfo {author} {\bibfnamefont {D.}~\bibnamefont {Gerace}},
  \bibinfo {author} {\bibfnamefont {M.}~\bibnamefont {Campisi}}, \ and\
  \bibinfo {author} {\bibfnamefont {M.~F.}\ \bibnamefont {Santos}},\ }\href
  {\doibase 10.1103/physreva.103.062225} {\bibfield  {journal} {\bibinfo
  {journal} {Physical Review A}\ }\textbf {\bibinfo {volume} {103}} (\bibinfo
  {year} {2021}),\ 10.1103/physreva.103.062225}\BibitemShut {NoStop}%
\bibitem [{\citenamefont {Zhao}(2021)}]{zhao2021stochastic}%
  \BibitemOpen
  \bibfield  {author} {\bibinfo {author} {\bibfnamefont {D.}~\bibnamefont
  {Zhao}},\ }\href@noop {} {\enquote {\bibinfo {title} {A stochastic heat
  engine based on prandtl-tomlinson model},}\ } (\bibinfo {year} {2021}),\
  \Eprint {http://arxiv.org/abs/2112.12536} {arXiv:2112.12536
  [physics.class-ph]} \BibitemShut {NoStop}%
\bibitem [{\citenamefont {Fu}\ \emph {et~al.}(2022)\citenamefont {Fu},
  \citenamefont {Miangolarra}, \citenamefont {Taghvaei}, \citenamefont {Chen},\
  and\ \citenamefont {Georgiou}}]{fu2022stochastic}%
  \BibitemOpen
  \bibfield  {author} {\bibinfo {author} {\bibfnamefont {R.}~\bibnamefont
  {Fu}}, \bibinfo {author} {\bibfnamefont {O.~M.}\ \bibnamefont {Miangolarra}},
  \bibinfo {author} {\bibfnamefont {A.}~\bibnamefont {Taghvaei}}, \bibinfo
  {author} {\bibfnamefont {Y.}~\bibnamefont {Chen}}, \ and\ \bibinfo {author}
  {\bibfnamefont {T.~T.}\ \bibnamefont {Georgiou}},\ }\href@noop {} {\enquote
  {\bibinfo {title} {Stochastic thermodynamic engines under time-varying
  temperature profile},}\ } (\bibinfo {year} {2022}),\ \Eprint
  {http://arxiv.org/abs/2207.05069} {arXiv:2207.05069 [cond-mat.stat-mech]}
  \BibitemShut {NoStop}%
\bibitem [{\citenamefont {Kumari}\ \emph {et~al.}(2023)\citenamefont {Kumari},
  \citenamefont {Samsuzzaman}, \citenamefont {Saha},\ and\ \citenamefont
  {Lahiri}}]{kumari2023stochastic}%
  \BibitemOpen
  \bibfield  {author} {\bibinfo {author} {\bibfnamefont {A.}~\bibnamefont
  {Kumari}}, \bibinfo {author} {\bibfnamefont {M.}~\bibnamefont {Samsuzzaman}},
  \bibinfo {author} {\bibfnamefont {A.}~\bibnamefont {Saha}}, \ and\ \bibinfo
  {author} {\bibfnamefont {S.}~\bibnamefont {Lahiri}},\ }\href@noop {}
  {\enquote {\bibinfo {title} {Stochastic heat engine using multiple
  interacting active particles},}\ } (\bibinfo {year} {2023}),\ \Eprint
  {http://arxiv.org/abs/2304.11867} {arXiv:2304.11867 [cond-mat.stat-mech]}
  \BibitemShut {NoStop}%
\bibitem [{\citenamefont {Ge}\ \emph {et~al.}(2012)\citenamefont {Ge},
  \citenamefont {Qian},\ and\ \citenamefont {Qian}}]{ge2012stochastic}%
  \BibitemOpen
  \bibfield  {author} {\bibinfo {author} {\bibfnamefont {H.}~\bibnamefont
  {Ge}}, \bibinfo {author} {\bibfnamefont {M.}~\bibnamefont {Qian}}, \ and\
  \bibinfo {author} {\bibfnamefont {H.}~\bibnamefont {Qian}},\ }\href@noop {}
  {\bibfield  {journal} {\bibinfo  {journal} {Physics Reports}\ }\textbf
  {\bibinfo {volume} {510}},\ \bibinfo {pages} {87} (\bibinfo {year}
  {2012})}\BibitemShut {NoStop}%
\bibitem [{\citenamefont {Liepelt}\ and\ \citenamefont
  {Lipowsky}(2007)}]{liepelt1}%
  \BibitemOpen
  \bibfield  {author} {\bibinfo {author} {\bibfnamefont {S.}~\bibnamefont
  {Liepelt}}\ and\ \bibinfo {author} {\bibfnamefont {R.}~\bibnamefont
  {Lipowsky}},\ }\href {\doibase 10.1103/PhysRevLett.98.258102} {\bibfield
  {journal} {\bibinfo  {journal} {Phys. Rev. Lett.}\ }\textbf {\bibinfo
  {volume} {98}},\ \bibinfo {pages} {258102} (\bibinfo {year}
  {2007})}\BibitemShut {NoStop}%
\bibitem [{\citenamefont {Liepelt}\ and\ \citenamefont
  {Lipowsky}(2009)}]{liepelt2}%
  \BibitemOpen
  \bibfield  {author} {\bibinfo {author} {\bibfnamefont {S.}~\bibnamefont
  {Liepelt}}\ and\ \bibinfo {author} {\bibfnamefont {R.}~\bibnamefont
  {Lipowsky}},\ }\href {\doibase 10.1103/PhysRevE.79.011917} {\bibfield
  {journal} {\bibinfo  {journal} {Phys. Rev. E}\ }\textbf {\bibinfo {volume}
  {79}},\ \bibinfo {pages} {011917} (\bibinfo {year} {2009})}\BibitemShut
  {NoStop}%
\bibitem [{\citenamefont {Busiello}\ and\ \citenamefont
  {Fiore}(2022)}]{busiello2022hyperaccurate}%
  \BibitemOpen
  \bibfield  {author} {\bibinfo {author} {\bibfnamefont {D.~M.}\ \bibnamefont
  {Busiello}}\ and\ \bibinfo {author} {\bibfnamefont {C.}~\bibnamefont
  {Fiore}},\ }\href@noop {} {\bibfield  {journal} {\bibinfo  {journal} {Journal
  of Physics A: Mathematical and Theoretical}\ }\textbf {\bibinfo {volume}
  {55}},\ \bibinfo {pages} {485004} (\bibinfo {year} {2022})}\BibitemShut
  {NoStop}%
\bibitem [{\citenamefont {Mamede}\ \emph {et~al.}(2023)\citenamefont {Mamede},
  \citenamefont {Proesmans},\ and\ \citenamefont
  {Fiore}}]{mamede2023thermodynamics}%
  \BibitemOpen
  \bibfield  {author} {\bibinfo {author} {\bibfnamefont {I.~N.}\ \bibnamefont
  {Mamede}}, \bibinfo {author} {\bibfnamefont {K.}~\bibnamefont {Proesmans}}, \
  and\ \bibinfo {author} {\bibfnamefont {C.~E.}\ \bibnamefont {Fiore}},\
  }\href@noop {} {\bibfield  {journal} {\bibinfo  {journal} {arXiv preprint
  arXiv:2308.02255}\ } (\bibinfo {year} {2023})}\BibitemShut {NoStop}%
\bibitem [{\citenamefont {Vroylandt}\ \emph {et~al.}(2017)\citenamefont
  {Vroylandt}, \citenamefont {Esposito},\ and\ \citenamefont
  {Verley}}]{gatien}%
  \BibitemOpen
  \bibfield  {author} {\bibinfo {author} {\bibfnamefont {H.}~\bibnamefont
  {Vroylandt}}, \bibinfo {author} {\bibfnamefont {M.}~\bibnamefont {Esposito}},
  \ and\ \bibinfo {author} {\bibfnamefont {G.}~\bibnamefont {Verley}},\ }\href
  {\doibase 10.1209/0295-5075/120/30009} {\bibfield  {journal} {\bibinfo
  {journal} {{EPL} (Europhysics Letters)}\ }\textbf {\bibinfo {volume} {120}},\
  \bibinfo {pages} {30009} (\bibinfo {year} {2017})}\BibitemShut {NoStop}%
\bibitem [{\citenamefont {Filho}\ \emph {et~al.}(2023)\citenamefont {Filho},
  \citenamefont {For\~ao}, \citenamefont {Busiello}, \citenamefont {Cleuren},\
  and\ \citenamefont {Fiore}}]{forao2023powerful}%
  \BibitemOpen
  \bibfield  {author} {\bibinfo {author} {\bibfnamefont {F.~S.}\ \bibnamefont
  {Filho}}, \bibinfo {author} {\bibfnamefont {G.~A.~L.}\ \bibnamefont
  {For\~ao}}, \bibinfo {author} {\bibfnamefont {D.~M.}\ \bibnamefont
  {Busiello}}, \bibinfo {author} {\bibfnamefont {B.}~\bibnamefont {Cleuren}}, \
  and\ \bibinfo {author} {\bibfnamefont {C.~E.}\ \bibnamefont {Fiore}},\ }\href
  {\doibase 10.1103/PhysRevResearch.5.043067} {\bibfield  {journal} {\bibinfo
  {journal} {Phys. Rev. Res.}\ }\textbf {\bibinfo {volume} {5}},\ \bibinfo
  {pages} {043067} (\bibinfo {year} {2023})}\BibitemShut {NoStop}%
\bibitem [{\citenamefont {Proesmans}\ \emph {et~al.}(2015)\citenamefont
  {Proesmans}, \citenamefont {Driesen}, \citenamefont {Cleuren},\ and\
  \citenamefont {Van~den Broeck}}]{proesmans2015efficiency}%
  \BibitemOpen
  \bibfield  {author} {\bibinfo {author} {\bibfnamefont {K.}~\bibnamefont
  {Proesmans}}, \bibinfo {author} {\bibfnamefont {C.}~\bibnamefont {Driesen}},
  \bibinfo {author} {\bibfnamefont {B.}~\bibnamefont {Cleuren}}, \ and\
  \bibinfo {author} {\bibfnamefont {C.}~\bibnamefont {Van~den Broeck}},\
  }\href@noop {} {\bibfield  {journal} {\bibinfo  {journal} {Physical review
  E}\ }\textbf {\bibinfo {volume} {92}},\ \bibinfo {pages} {032105} (\bibinfo
  {year} {2015})}\BibitemShut {NoStop}%
\bibitem [{\citenamefont {Proesmans}\ \emph {et~al.}(2016)\citenamefont
  {Proesmans}, \citenamefont {Dreher}, \citenamefont {Gavrilov}, \citenamefont
  {Bechhoefer},\ and\ \citenamefont {Van~den Broeck}}]{proesmans2016brownian}%
  \BibitemOpen
  \bibfield  {author} {\bibinfo {author} {\bibfnamefont {K.}~\bibnamefont
  {Proesmans}}, \bibinfo {author} {\bibfnamefont {Y.}~\bibnamefont {Dreher}},
  \bibinfo {author} {\bibfnamefont {M.}~\bibnamefont {Gavrilov}}, \bibinfo
  {author} {\bibfnamefont {J.}~\bibnamefont {Bechhoefer}}, \ and\ \bibinfo
  {author} {\bibfnamefont {C.}~\bibnamefont {Van~den Broeck}},\ }\href@noop {}
  {\bibfield  {journal} {\bibinfo  {journal} {Physical Review X}\ }\textbf
  {\bibinfo {volume} {6}},\ \bibinfo {pages} {041010} (\bibinfo {year}
  {2016})}\BibitemShut {NoStop}%
\bibitem [{\citenamefont {Proesmans}\ and\ \citenamefont {Van~den
  Broeck}(2017)}]{proesmans2017underdamped}%
  \BibitemOpen
  \bibfield  {author} {\bibinfo {author} {\bibfnamefont {K.}~\bibnamefont
  {Proesmans}}\ and\ \bibinfo {author} {\bibfnamefont {C.}~\bibnamefont
  {Van~den Broeck}},\ }\href@noop {} {\bibfield  {journal} {\bibinfo  {journal}
  {Chaos: An Interdisciplinary Journal of Nonlinear Science}\ }\textbf
  {\bibinfo {volume} {27}},\ \bibinfo {pages} {104601} (\bibinfo {year}
  {2017})}\BibitemShut {NoStop}%
\bibitem [{\citenamefont {Mamede}\ \emph
  {et~al.}(2022{\natexlab{a}})\citenamefont {Mamede}, \citenamefont {Harunari},
  \citenamefont {Akasaki}, \citenamefont {Proesmans},\ and\ \citenamefont
  {Fiore}}]{mamede2021obtaining}%
  \BibitemOpen
  \bibfield  {author} {\bibinfo {author} {\bibfnamefont {I.~N.}\ \bibnamefont
  {Mamede}}, \bibinfo {author} {\bibfnamefont {P.~E.}\ \bibnamefont
  {Harunari}}, \bibinfo {author} {\bibfnamefont {B.~A.~N.}\ \bibnamefont
  {Akasaki}}, \bibinfo {author} {\bibfnamefont {K.}~\bibnamefont {Proesmans}},
  \ and\ \bibinfo {author} {\bibfnamefont {C.~E.}\ \bibnamefont {Fiore}},\
  }\href {\doibase 10.1103/PhysRevE.105.024106} {\bibfield  {journal} {\bibinfo
   {journal} {Phys. Rev. E}\ }\textbf {\bibinfo {volume} {105}},\ \bibinfo
  {pages} {024106} (\bibinfo {year} {2022}{\natexlab{a}})}\BibitemShut
  {NoStop}%
\bibitem [{\citenamefont {Noa}\ \emph {et~al.}(2020)\citenamefont {Noa},
  \citenamefont {Oropesa},\ and\ \citenamefont
  {Fiore}}]{noa2020thermodynamics}%
  \BibitemOpen
  \bibfield  {author} {\bibinfo {author} {\bibfnamefont {C.~F.}\ \bibnamefont
  {Noa}}, \bibinfo {author} {\bibfnamefont {W.~G.}\ \bibnamefont {Oropesa}}, \
  and\ \bibinfo {author} {\bibfnamefont {C.}~\bibnamefont {Fiore}},\
  }\href@noop {} {\bibfield  {journal} {\bibinfo  {journal} {Physical Review
  Research}\ }\textbf {\bibinfo {volume} {2}},\ \bibinfo {pages} {043016}
  (\bibinfo {year} {2020})}\BibitemShut {NoStop}%
\bibitem [{\citenamefont {Noa}\ \emph {et~al.}(2021)\citenamefont {Noa},
  \citenamefont {Stable}, \citenamefont {Oropesa}, \citenamefont {Rosas},\ and\
  \citenamefont {Fiore}}]{noa2021efficient}%
  \BibitemOpen
  \bibfield  {author} {\bibinfo {author} {\bibfnamefont {C.~E.~F.}\
  \bibnamefont {Noa}}, \bibinfo {author} {\bibfnamefont {A.~L.~L.}\
  \bibnamefont {Stable}}, \bibinfo {author} {\bibfnamefont {W.~G.~C.}\
  \bibnamefont {Oropesa}}, \bibinfo {author} {\bibfnamefont {A.}~\bibnamefont
  {Rosas}}, \ and\ \bibinfo {author} {\bibfnamefont {C.~E.}\ \bibnamefont
  {Fiore}},\ }\href {\doibase 10.1103/PhysRevResearch.3.043152} {\bibfield
  {journal} {\bibinfo  {journal} {Phys. Rev. Research}\ }\textbf {\bibinfo
  {volume} {3}},\ \bibinfo {pages} {043152} (\bibinfo {year}
  {2021})}\BibitemShut {NoStop}%
\bibitem [{\citenamefont {Mamede}\ \emph
  {et~al.}(2022{\natexlab{b}})\citenamefont {Mamede}, \citenamefont {Stable},\
  and\ \citenamefont {Fiore}}]{mamede2022}%
  \BibitemOpen
  \bibfield  {author} {\bibinfo {author} {\bibfnamefont {I.~N.}\ \bibnamefont
  {Mamede}}, \bibinfo {author} {\bibfnamefont {A.~L.~L.}\ \bibnamefont
  {Stable}}, \ and\ \bibinfo {author} {\bibfnamefont {C.~E.}\ \bibnamefont
  {Fiore}},\ }\href {\doibase 10.1103/PhysRevE.106.064125} {\bibfield
  {journal} {\bibinfo  {journal} {Phys. Rev. E}\ }\textbf {\bibinfo {volume}
  {106}},\ \bibinfo {pages} {064125} (\bibinfo {year}
  {2022}{\natexlab{b}})}\BibitemShut {NoStop}%
\bibitem [{\citenamefont {Rosas}\ \emph {et~al.}(2017)\citenamefont {Rosas},
  \citenamefont {Van~den Broeck},\ and\ \citenamefont {Lindenberg}}]{rosas1}%
  \BibitemOpen
  \bibfield  {author} {\bibinfo {author} {\bibfnamefont {A.}~\bibnamefont
  {Rosas}}, \bibinfo {author} {\bibfnamefont {C.}~\bibnamefont {Van~den
  Broeck}}, \ and\ \bibinfo {author} {\bibfnamefont {K.}~\bibnamefont
  {Lindenberg}},\ }\href {\doibase 10.1103/PhysRevE.96.052135} {\bibfield
  {journal} {\bibinfo  {journal} {Phys. Rev. E}\ }\textbf {\bibinfo {volume}
  {96}},\ \bibinfo {pages} {052135} (\bibinfo {year} {2017})}\BibitemShut
  {NoStop}%
\bibitem [{\citenamefont {Rosas}\ \emph {et~al.}(2018)\citenamefont {Rosas},
  \citenamefont {Van~den Broeck},\ and\ \citenamefont {Lindenberg}}]{rosas2}%
  \BibitemOpen
  \bibfield  {author} {\bibinfo {author} {\bibfnamefont {A.}~\bibnamefont
  {Rosas}}, \bibinfo {author} {\bibfnamefont {C.}~\bibnamefont {Van~den
  Broeck}}, \ and\ \bibinfo {author} {\bibfnamefont {K.}~\bibnamefont
  {Lindenberg}},\ }\href {\doibase 10.1103/PhysRevE.97.062103} {\bibfield
  {journal} {\bibinfo  {journal} {Phys. Rev. E}\ }\textbf {\bibinfo {volume}
  {97}},\ \bibinfo {pages} {062103} (\bibinfo {year} {2018})}\BibitemShut
  {NoStop}%
\bibitem [{\citenamefont {Ray}\ and\ \citenamefont
  {Barato}(2017)}]{barato2017}%
  \BibitemOpen
  \bibfield  {author} {\bibinfo {author} {\bibfnamefont {S.}~\bibnamefont
  {Ray}}\ and\ \bibinfo {author} {\bibfnamefont {A.~C.}\ \bibnamefont
  {Barato}},\ }\href {\doibase 10.1103/PhysRevE.96.052120} {\bibfield
  {journal} {\bibinfo  {journal} {Phys. Rev. E}\ }\textbf {\bibinfo {volume}
  {96}},\ \bibinfo {pages} {052120} (\bibinfo {year} {2017})}\BibitemShut
  {NoStop}%
\bibitem [{\citenamefont {Bennett}(1982)}]{bennett1982thermodynamics}%
  \BibitemOpen
  \bibfield  {author} {\bibinfo {author} {\bibfnamefont {C.~H.}\ \bibnamefont
  {Bennett}},\ }\href@noop {} {\bibfield  {journal} {\bibinfo  {journal}
  {International Journal of Theoretical Physics}\ }\textbf {\bibinfo {volume}
  {21}},\ \bibinfo {pages} {905} (\bibinfo {year} {1982})}\BibitemShut
  {NoStop}%
\bibitem [{\citenamefont {Sagawa}(2014)}]{sagawa2014thermodynamic}%
  \BibitemOpen
  \bibfield  {author} {\bibinfo {author} {\bibfnamefont {T.}~\bibnamefont
  {Sagawa}},\ }\href@noop {} {\bibfield  {journal} {\bibinfo  {journal}
  {Journal of Statistical Mechanics: Theory and Experiment}\ }\textbf {\bibinfo
  {volume} {2014}},\ \bibinfo {pages} {P03025} (\bibinfo {year}
  {2014})}\BibitemShut {NoStop}%
\bibitem [{\citenamefont {Giovannetti}\ and\ \citenamefont
  {Palma}(2012)}]{PhysRevLett.108.040401}%
  \BibitemOpen
  \bibfield  {author} {\bibinfo {author} {\bibfnamefont {V.}~\bibnamefont
  {Giovannetti}}\ and\ \bibinfo {author} {\bibfnamefont {G.~M.}\ \bibnamefont
  {Palma}},\ }\href {\doibase 10.1103/PhysRevLett.108.040401} {\bibfield
  {journal} {\bibinfo  {journal} {Phys. Rev. Lett.}\ }\textbf {\bibinfo
  {volume} {108}},\ \bibinfo {pages} {040401} (\bibinfo {year}
  {2012})}\BibitemShut {NoStop}%
\bibitem [{\citenamefont {Harunari}\ \emph {et~al.}(2021)\citenamefont
  {Harunari}, \citenamefont {Filho}, \citenamefont {Fiore},\ and\ \citenamefont
  {Rosas}}]{harunari2021maximal}%
  \BibitemOpen
  \bibfield  {author} {\bibinfo {author} {\bibfnamefont {P.~E.}\ \bibnamefont
  {Harunari}}, \bibinfo {author} {\bibfnamefont {F.~S.}\ \bibnamefont {Filho}},
  \bibinfo {author} {\bibfnamefont {C.~E.}\ \bibnamefont {Fiore}}, \ and\
  \bibinfo {author} {\bibfnamefont {A.}~\bibnamefont {Rosas}},\ }\href
  {\doibase 10.1103/PhysRevResearch.3.023194} {\bibfield  {journal} {\bibinfo
  {journal} {Phys. Rev. Research}\ }\textbf {\bibinfo {volume} {3}},\ \bibinfo
  {pages} {023194} (\bibinfo {year} {2021})}\BibitemShut {NoStop}%
\bibitem [{\citenamefont {Herpich}\ \emph {et~al.}(2018)\citenamefont
  {Herpich}, \citenamefont {Thingna},\ and\ \citenamefont
  {Esposito}}]{herpich}%
  \BibitemOpen
  \bibfield  {author} {\bibinfo {author} {\bibfnamefont {T.}~\bibnamefont
  {Herpich}}, \bibinfo {author} {\bibfnamefont {J.}~\bibnamefont {Thingna}}, \
  and\ \bibinfo {author} {\bibfnamefont {M.}~\bibnamefont {Esposito}},\ }\href
  {\doibase 10.1103/PhysRevX.8.031056} {\bibfield  {journal} {\bibinfo
  {journal} {Phys. Rev. X}\ }\textbf {\bibinfo {volume} {8}},\ \bibinfo {pages}
  {031056} (\bibinfo {year} {2018})}\BibitemShut {NoStop}%
\bibitem [{\citenamefont {Herpich}\ and\ \citenamefont
  {Esposito}(2019)}]{herpich2}%
  \BibitemOpen
  \bibfield  {author} {\bibinfo {author} {\bibfnamefont {T.}~\bibnamefont
  {Herpich}}\ and\ \bibinfo {author} {\bibfnamefont {M.}~\bibnamefont
  {Esposito}},\ }\href {\doibase 10.1103/PhysRevE.99.022135} {\bibfield
  {journal} {\bibinfo  {journal} {Phys. Rev. E}\ }\textbf {\bibinfo {volume}
  {99}},\ \bibinfo {pages} {022135} (\bibinfo {year} {2019})}\BibitemShut
  {NoStop}%
\bibitem [{\citenamefont {Bettmann}\ \emph {et~al.}(2023)\citenamefont
  {Bettmann}, \citenamefont {Kewming},\ and\ \citenamefont
  {Goold}}]{PhysRevE.107.044102}%
  \BibitemOpen
  \bibfield  {author} {\bibinfo {author} {\bibfnamefont {L.~P.}\ \bibnamefont
  {Bettmann}}, \bibinfo {author} {\bibfnamefont {M.~J.}\ \bibnamefont
  {Kewming}}, \ and\ \bibinfo {author} {\bibfnamefont {J.}~\bibnamefont
  {Goold}},\ }\href {\doibase 10.1103/PhysRevE.107.044102} {\bibfield
  {journal} {\bibinfo  {journal} {Phys. Rev. E}\ }\textbf {\bibinfo {volume}
  {107}},\ \bibinfo {pages} {044102} (\bibinfo {year} {2023})}\BibitemShut
  {NoStop}%
\bibitem [{\citenamefont {Prech}\ \emph {et~al.}(2023)\citenamefont {Prech},
  \citenamefont {Johansson}, \citenamefont {Nyholm}, \citenamefont {Landi},
  \citenamefont {Verdozzi}, \citenamefont {Samuelsson},\ and\ \citenamefont
  {Potts}}]{PhysRevResearch.5.023155}%
  \BibitemOpen
  \bibfield  {author} {\bibinfo {author} {\bibfnamefont {K.}~\bibnamefont
  {Prech}}, \bibinfo {author} {\bibfnamefont {P.}~\bibnamefont {Johansson}},
  \bibinfo {author} {\bibfnamefont {E.}~\bibnamefont {Nyholm}}, \bibinfo
  {author} {\bibfnamefont {G.~T.}\ \bibnamefont {Landi}}, \bibinfo {author}
  {\bibfnamefont {C.}~\bibnamefont {Verdozzi}}, \bibinfo {author}
  {\bibfnamefont {P.}~\bibnamefont {Samuelsson}}, \ and\ \bibinfo {author}
  {\bibfnamefont {P.~P.}\ \bibnamefont {Potts}},\ }\href {\doibase
  10.1103/PhysRevResearch.5.023155} {\bibfield  {journal} {\bibinfo  {journal}
  {Phys. Rev. Res.}\ }\textbf {\bibinfo {volume} {5}},\ \bibinfo {pages}
  {023155} (\bibinfo {year} {2023})}\BibitemShut {NoStop}%
\bibitem [{\citenamefont {Vroylandt}\ \emph {et~al.}(2020)\citenamefont
  {Vroylandt}, \citenamefont {Proesmans},\ and\ \citenamefont
  {Gingrich}}]{vroylandt2020isometric}%
  \BibitemOpen
  \bibfield  {author} {\bibinfo {author} {\bibfnamefont {H.}~\bibnamefont
  {Vroylandt}}, \bibinfo {author} {\bibfnamefont {K.}~\bibnamefont
  {Proesmans}}, \ and\ \bibinfo {author} {\bibfnamefont {T.~R.}\ \bibnamefont
  {Gingrich}},\ }\href@noop {} {\bibfield  {journal} {\bibinfo  {journal}
  {Journal of Statistical Physics}\ }\textbf {\bibinfo {volume} {178}},\
  \bibinfo {pages} {1039} (\bibinfo {year} {2020})}\BibitemShut {NoStop}%
\bibitem [{\citenamefont {Proesmans}\ and\ \citenamefont
  {Fiore}(2019)}]{fiorek}%
  \BibitemOpen
  \bibfield  {author} {\bibinfo {author} {\bibfnamefont {K.}~\bibnamefont
  {Proesmans}}\ and\ \bibinfo {author} {\bibfnamefont {C.~E.}\ \bibnamefont
  {Fiore}},\ }\href@noop {} {\bibfield  {journal} {\bibinfo  {journal}
  {Physical Review E}\ }\textbf {\bibinfo {volume} {100}},\ \bibinfo {pages}
  {022141} (\bibinfo {year} {2019})}\BibitemShut {NoStop}%
\bibitem [{\citenamefont {Tom{\'e}}\ and\ \citenamefont
  {de~Oliveira}(2015)}]{tome2015}%
  \BibitemOpen
  \bibfield  {author} {\bibinfo {author} {\bibfnamefont {T.}~\bibnamefont
  {Tom{\'e}}}\ and\ \bibinfo {author} {\bibfnamefont {M.~J.}\ \bibnamefont
  {de~Oliveira}},\ }\href
  {https://journals.aps.org/pre/abstract/10.1103/PhysRevE.91.042140} {\bibfield
   {journal} {\bibinfo  {journal} {Physical review E}\ }\textbf {\bibinfo
  {volume} {91}},\ \bibinfo {pages} {042140} (\bibinfo {year}
  {2015})}\BibitemShut {NoStop}%
\bibitem [{\citenamefont {Fernandez~Noa}\ \emph {et~al.}(2023)\citenamefont
  {Fernandez~Noa}, \citenamefont {Fiore}, \citenamefont {Wijns}, \citenamefont
  {Cleuren} \emph {et~al.}}]{noa2023thermodynamics}%
  \BibitemOpen
  \bibfield  {author} {\bibinfo {author} {\bibfnamefont {C.}~\bibnamefont
  {Fernandez~Noa}}, \bibinfo {author} {\bibfnamefont {C.}~\bibnamefont
  {Fiore}}, \bibinfo {author} {\bibfnamefont {B.}~\bibnamefont {Wijns}},
  \bibinfo {author} {\bibfnamefont {B.}~\bibnamefont {Cleuren}},  \emph
  {et~al.},\ }\href@noop {} {\bibfield  {journal} {\bibinfo  {journal} {arXiv
  e-prints}\ ,\ \bibinfo {pages} {arXiv}} (\bibinfo {year} {2023})}\BibitemShut
  {NoStop}%
\bibitem [{\citenamefont {Curzon}\ and\ \citenamefont
  {Ahlborn}(1975)}]{curzon1975efficiency}%
  \BibitemOpen
  \bibfield  {author} {\bibinfo {author} {\bibfnamefont {F.}~\bibnamefont
  {Curzon}}\ and\ \bibinfo {author} {\bibfnamefont {B.}~\bibnamefont
  {Ahlborn}},\ }\href@noop {} {\bibfield  {journal} {\bibinfo  {journal}
  {American Journal of Physics}\ }\textbf {\bibinfo {volume} {43}},\ \bibinfo
  {pages} {22} (\bibinfo {year} {1975})}\BibitemShut {NoStop}%
\bibitem [{\citenamefont {Novikov}(1958)}]{novikov1958efficiency}%
  \BibitemOpen
  \bibfield  {author} {\bibinfo {author} {\bibfnamefont {I.}~\bibnamefont
  {Novikov}},\ }\href@noop {} {\bibfield  {journal} {\bibinfo  {journal}
  {Journal of Nuclear Energy (1954)}\ }\textbf {\bibinfo {volume} {7}},\
  \bibinfo {pages} {125} (\bibinfo {year} {1958})}\BibitemShut {NoStop}%
\end{thebibliography}%
\end{document}